\documentclass[epj]{svjour_mod}
\usepackage[latin9]{inputenc}
\usepackage[T1]{fontenc}
\usepackage{longtable}
\usepackage{ a4}
\usepackage{float}
\usepackage{graphicx}
\usepackage{epstopdf}
\usepackage{subfigure}
\usepackage{amssymb}
\usepackage{fancybox}
\usepackage{amsmath}
\usepackage{epsfig}
\usepackage{dsfont}
\usepackage{fancyhdr}
\usepackage{picins}

\begin{document}
\include{00README.XXX}
\title{Determination of the extraction efficiency for $^{233}$U source $\alpha$-recoil ions from the MLL buffer-gas stopping cell}
\subtitle{}
\author{Lars v.d.Wense\inst{1} \and Benedict Seiferle\inst{1} 
        \and Mustapha Laatiaoui\inst{2,3} \and Peter G. Thirolf\inst{1}
}                     
%
%
\institute{Ludwig-Maximilians-Universit\"at M\"unchen, Am Coulombwall 1, 
           Garching, Germany 
\and 
        GSI Helmholtzzentrum f\"ur Schwerionenforschung GmbH, Planckstr. 1, 
        Darmstadt, Germany 
\and 
        Helmholtz Institut Mainz, Johann-Joachim-Becherweg 36, Mainz, Germany }
\date{6. March 2015}
\abstract{
Following the $\alpha$ decay of $^{233}$U, $^{229}$Th recoil ions are shown to be extracted in a significant amount from the MLL buffer-gas stopping cell. The produced recoil ions and subsequent daughter nuclei are mass purified with the help of a customized quadrupole mass spectrometer. The combined extraction and mass-purification efficiency for $^{229}$Th$^{3+}$ is determined via MCP-based measurements and via the direct detection of the $^{229}$Th $\alpha$ decay. A large value of $(10\pm2)$\% for the combined extraction and mass-purification efficiency of $^{229}$Th$^{3+}$ is obtained at a mass resolution of about 1 u/e. In addition to $^{229}$Th, also other $\alpha$-recoil ions of the $^{233,232}$U decay chains are addressed. 
} 
\titlerunning{Extraction efficiencies for $^{233}$U source $\alpha$-recoil ions}
\maketitle
\section{Introduction}
\label{intro}
Since it was proposed that the isomeric lowest excited nuclear state of $^{229}$Th could 
have an interesting application as a highly stable nuclear-based frequency 
standard \cite{Peik}, there has been an increasing interest in the direct detection of 
the corresponding ground-state transition and a precise measurement of its energy.

$^{229}$Th was first considered already in 1976 by Kroger and Reich \cite{Kroger_Reich} 
to possess a low-lying nuclear excitation. The energy was then inferred to be $3.5\pm1$ 
eV \cite{Helmer_Reich} and more recently corrected to $7.6\pm0.5$ eV 
($163\pm11$ nm)~\cite{Beck}. Both measurements were based on double differences of 
higher-lying rotational $\gamma$ transitions and have to be considered as indirect 
measurements. The measured energy is very low compared to energies typically involved 
in nuclear processes and even matches typical energies of atomic shell transitions. 
Actually it is the lowest nuclear excitation presently known in the landscape of atomic 
nuclei. Therefore, it potentially allows for the application of laser spectroscopic 
methods. The predicted value for the half-life significantly depends on the energy of 
the isomeric transition and the electronic surrounding, as internal conversion (IC) 
effects and bound internal conversion (BIC) effects are expected to play a significant 
role. If internal conversion is allowed by the electronic environment, the isomer is 
expected to predominantly decay via the IC channel ($\alpha_{IC}\approx 10^{9}$) and the 
half-life is predicted to be as short as $10^{-5}$ s \cite{Trzhaskovskaya1}. In case 
that IC is suppressed, but BIC has to be considered, life-times of minutes to hours have 
been predicted \cite{Trzhaskovskaya1,Trzhaskovskaya2,Tkalya1,Tkalya2}. A BIC decay 
enhancement factor of 400-600 compared to the radiative decay was estimated for a 3.5 eV isomeric transition energy \cite{Trzhaskovskaya2}.  Assuming an experimental setup, which allows for the suppression of the IC decay channel, the expected long half-life would cause the linewidth of the ground-state transition to be exceptionally small (of the order of $\Delta E/E \approx 10^{-20}$). These properties, combined with an expected high resistance against variations of external electric and magnetic fields \cite{Peik2}, render the isomeric transition a promising candidate for a nuclear-based frequency standard. However, in order to gain an optical access to the transition, the corresponding wavelength has to be known to higher precision. To ultimately prove the existence of the isomeric transition and to further improve the knowledge about its energy, a direct detection of the ground-state transition has been envisaged by several groups \cite{Swanberg,Rellergert,Porsev,Kazakov,Raeder}.

Most recently, a direct detection was claimed by X. Zhao {\it et al.} \cite{Zhao}, but the result was stated doubtful \cite{Peik3}. In the approach of Ref. \cite{Zhao}, $^{229}$Th recoils, as produced in the $\alpha$ decay of $^{233}$U, were implanted into a MgF$_2$ crystal. As there is a 2\% $\alpha$-decay branch into the isomeric state, the corresponding $\gamma$ decay fluorescence of $^{229m}$Th is expected to be detectable. There are several sources of background radiation discussed by X. Zhao {\it et al.}, one caused by the implantation of short-lived daughter nuclei from the $^{233}$U as well as $^{232}$U decay chains. $^{232}$U always has to be expected as a small impurity in $^{233}$U sources, caused by the radiochemical production process. Due to the shorter half-lives of  $^{232}$U (68.9 yr) and $^{228}$Th (1.9 yr) compared to $^{233}$U (1.592 $\cdot 10^5$ yr) and $^{229}$Th (7932 yr), respectively, the activities of the decay chains below thorium are of the same order of magnitude. In order to exclude background radiation in form of Cherenkov light and scintillation, the source used by X. Zhao {\it et al.} was chemically purified to remove more than 99.99\% of the uranium daughters. Measurements were performed with this source for up to 140 days after chemical purification. It has been argued in \cite{Peik3}, that the ingrowth of the daughters into this source may have been large enough to still lead to a significant background radiation and therefore may have diluted the radiation caused by the isomeric decay.

Background radiation caused by the short-lived isotopes of the decay chains of $^{233}$U and $^{232}$U, respectively, has so far been a problem in all approaches of searching for the isomeric transition following the $\alpha$ decay of $^{233}$U \cite{Zimmermann,Swanberg2}. Our experimental approach is aiming for a conceptual solution of this problem by allowing for a mass purification of the recoil ions leaving the uranium source. Therefore, the recoil ions are stopped with the help of a buffer-gas stopping cell. Subsequently, a low energy ion beam is formed, allowing for mass separation with the help of a quadrupole mass spectrometer. It is then envisaged to collect the mass purified ions on a 50 $\mu$m diameter, MgF$_2$ coated collection electrode. This coating is expected to support the radiative decay, as the effect of quenching should be suppressed. Further, thorium, when implanted into the MgF$_2$ crystal, should prefer the 4$^{+}$ charge state and therefore internal conversion is expected to be energetically forbidden. The isomeric $\gamma$ decay, if it exists, will take place from this point-like source and therefore will allow for applying highly efficient VUV optics, leading to a good signal to background ratio and open up the possibility of a more precise wavelength determination \cite{Wense}.

It is obvious that the extraction and mass-purification efficiency is of major interest for this experimental concept. Therefore, significant experimental effort was put in the determination and optimization of the ion extraction system. In the following section the experimental setup of the ion extraction system is described.

\section{The ion extraction system}
\label{sec:1}
\begin{figure*}
 \resizebox{1.0\textwidth}{!}{%
 \includegraphics{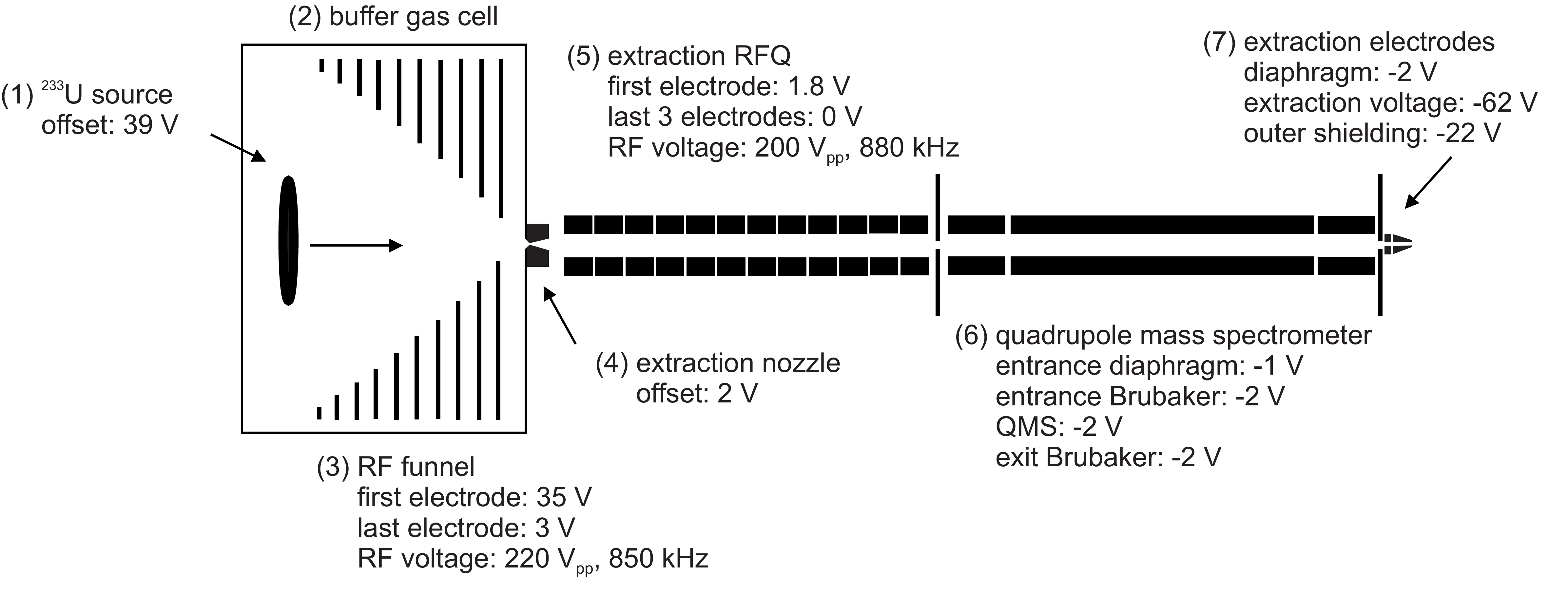}}
 \caption{Conceptual sketch of the ion extraction system designed for the highly efficient extraction and low-energy ion beam formation from a $^{233}$U $\alpha$-recoil ion source. The optimized extraction voltages are indicated for each electrode. The overall length of the system is about 85 cm. A description of the extraction system is given in the text.}
 \label{extraction_setup}
\end{figure*}

A conceptual sketch of the ion extraction system is shown in fig. \ref{extraction_setup}. The index numbers given in the text correspond to the numbers indicated in this sketch. The optimized extraction voltages are listed for each electrode.\\

In the described experimental approach, the isomer $^{229m}$Th is populated via a $2\%$ decay branch in the $\alpha$ decay of $^{233}$U. 200 kBq of $^{233}$U (UF$_4$), evaporated onto a 20 mm diameter aluminum plate, acts as an $\alpha$-recoil source (1). The source is placed inside a buffer-gas stopping cell (2), fully designed to ultra-high vacuum standards and bakeable up to 180$^\circ$ \cite{Neumayr}. After baking for 48 hours at 130$^\circ$,
 a typical pressure of 2.8$\cdot 10^{-10}$ mbar is reached. The cell is then flooded with 40 mbar of ultra-pure helium 6.0, which is further cleaned by catalytic purification (to the ppb level) and a cryo trap, filled with liquid nitrogen. The gas tubing was electropolished to achieve the highest cleanliness. All $\alpha$-decay recoil nuclei, which leave the source, are stopped within the gas. The stopping range for $^{229}$Th recoil ions in the helium gas was calculated using TRIM \cite{Ziegler} to be about 10 mm. The highly pure environment of the vacuum system is required to allow for a significant fraction of the recoil ions to survive the stopping and subsequent extraction as (multiply) charged ions. As will be shown later, the high cleanliness in our setup enabled the extraction of $^{229}$Th recoil ions even in the 3+ charge state.
 
 An electrode system inside the gas cell provides a guiding field, designed to drag the stopped ions fast and efficiently towards an extraction nozzle. The $^{233}$U source itself is put onto a voltage offset of 39 V, in order to push the ions towards an electric funnel system (3). The source is placed directly in front of the electric funnel in order to provide a short extraction path for the $\alpha$-recoil ions. The funnel system consists of 50 ring electrodes, converging in diameter towards the extraction nozzle. The electrodes are provided with a DC voltage gradient of 4 V/cm and RF voltages (220 V$_{\text{pp}}$, 850 kHz), alternating in phase by 180$^\circ$. 
The DC voltage guides the ions towards the chamber exit, while the RF voltages lead to a repelling force, preventing those ions, that are stopped off-axis, from discharging at the funnel electrodes.

The exit of the stopping cell is formed by a supersonic Laval nozzle with a 0.6 mm diameter nozzle throat and a voltage offset of 2 V (4). The helium gas is extracted from the cell through this nozzle, forming a supersonic gas jet when entering a second vacuum chamber with an ambient pressure of typically $2\cdot 10^{-2}$ mbar. When the gas jet is formed, the recoil ions are dragged away from the electric field lines and enter the extraction chamber together with the carrier gas. Here they are injected into a radio frequency quadrupole (RFQ) system (5), by which they are radially confined, resulting in the formation of an ion beam. The ambient helium pressure leads to a phase-space cooling of the beam, resulting in the formation of a sub-mm diameter recoil-ion beam at the RFQ exit. The extraction RFQ rod system is segmented into 12 parts, to each of which a separate DC voltage can be applied, in order to drag the ions through the helium background. The voltage gradient applied is 0.1 V/cm, where the last 3 rods are set to 0 V offset. The RF amplitude is 200 V$_{\text{pp}}$ at a frequency of 880 kHz for 10 mm inner rod distance. Within earlier measurements \cite{Neumayr2}, the extraction time from the buffer-gas stopping cell was determined to be in the range of a few ms (3-5 ms were obtained). Combined with the high cleanliness of the system, this fast extraction allows the ions to stay (multiply) charged. At the same time, isomeric half-lives well within the ms range can be probed.  

\begin{figure}
 \begin{center}
 \resizebox{0.5\textwidth}{!}{%
 \includegraphics{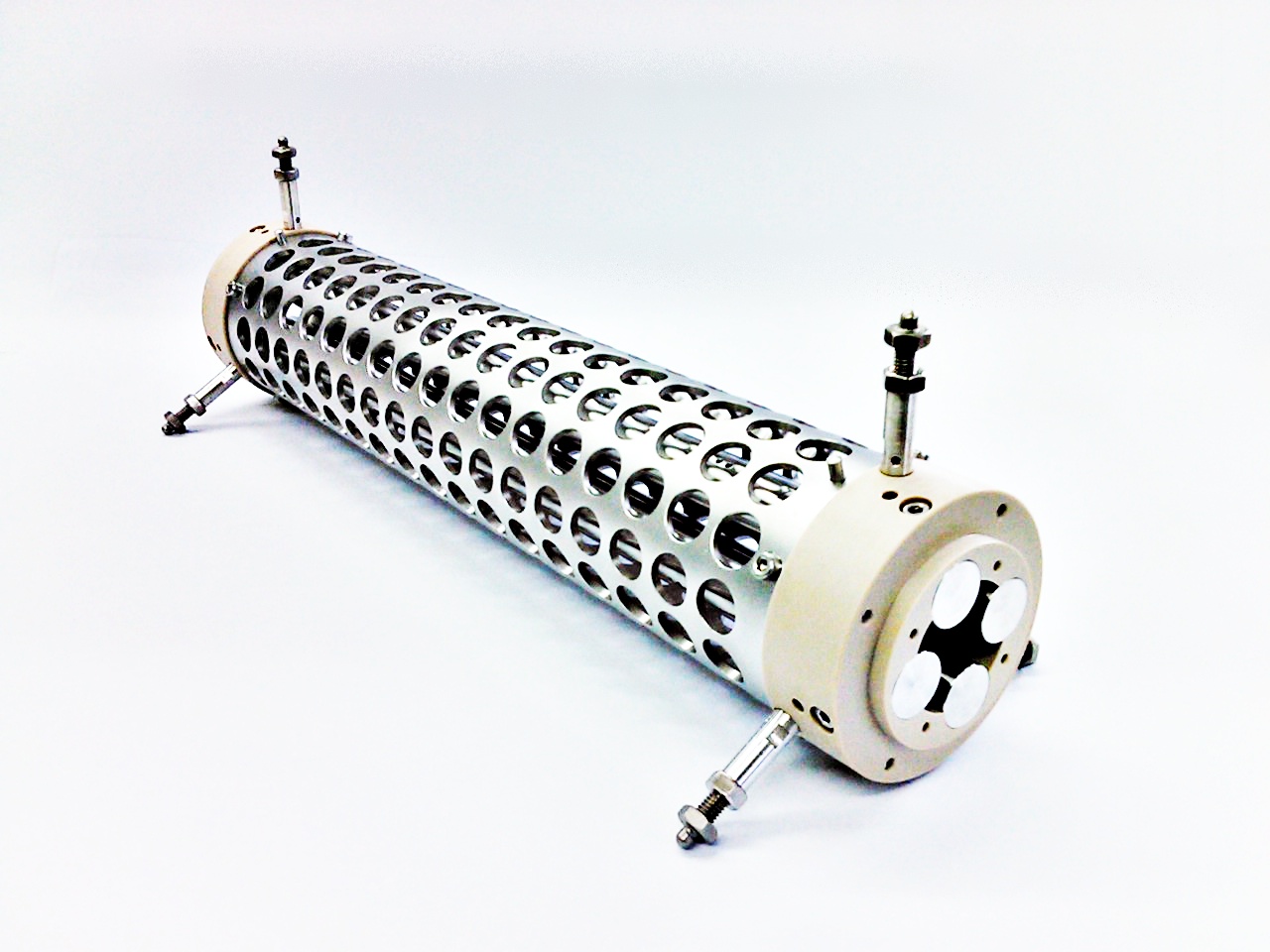}}
 \caption{Quadruple Mass Spectrometer as designed for a suppression of the short lived isotopes contained in the $^{232,233}$U decay chains. The overall length is about 40 cm.}
 \label{QMS_image}
 \end{center}
\end{figure}

\begin{figure}
 \begin{center}
 \resizebox{0.5\textwidth}{!}{%
 \includegraphics{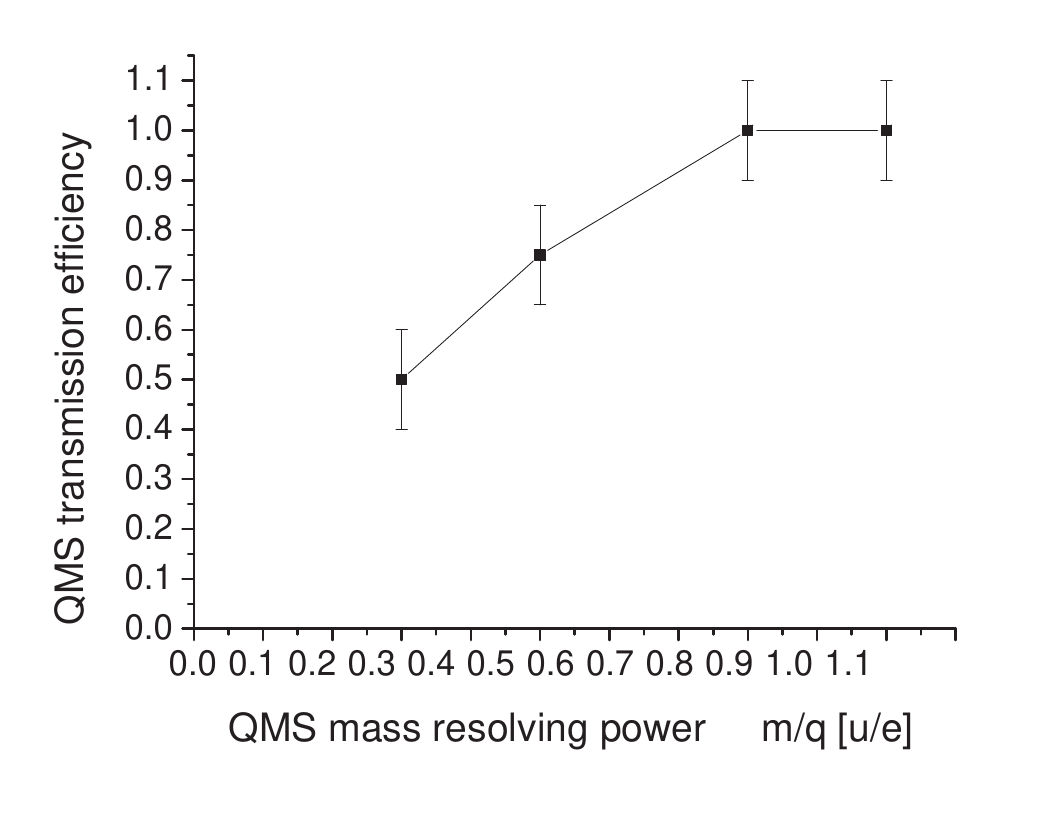}}
 \caption{QMS transmission efficiency as a function of the mass resolving power 
          as observed for the extraction of $^{229}$Th$^{3+}$.}
 \label{QMS_transmit}
 \end{center}
\end{figure}

At this point, all $\alpha$-recoil isotopes of the decay chains are extracted, however having formed an ion beam, mass separation may be realized with the help of a customized quadrupole mass spectrometer (QMS) (6) shown in fig. \ref{QMS_image}. The QMS was designed aiming at a maximum transmission efficiency, while suppressing the short-lived isotopes of the $^{232,233}$U decay chains. This includes all isotopes below $^{229}$Th and $^{228}$Th, respectively. The mass difference between neighbouring isotopes in the decay chains is 4 u, therefore, a mass resolving power of better than 2 u at mass 229 u was envisaged to safely allow for the suppression of the short-lived isotopes. This leads to the requirement of a mass resolving power of $m/\Delta m \geq 115$ at a high transmission efficiency. For this reason, a QMS was built with 18 mm rod diameter and inner rod distance of 15.96 mm. The length is 30 cm with further 5 cm long segments at the entrance and exit, acting as Brubaker lenses \cite{Brubaker}. Besides a 2 mm entrance diaphragm, which was set to -1 V, a voltage offset of -2 V for the entire system was applied. The design values were taken from \cite{Haettner}, where also a detailed analysis of the mass resolving power was made. Measured at 10\% of the peak maximum, a mass resolving power of $m/\Delta m =160$ was found for 90\% transmission efficiency. In our measurements for $^{229}$Th$^{3+}$ at $m/\Delta m=150$ (corresponding to $\Delta m/q=0.5$ u/e), more than 70\% transmission efficiency could be obtained, as can be seen in fig. \ref{QMS_transmit}, where the QMS transmission efficiency as a function of the mass resolving power is shown. Therefore our QMS fulfills the given requirements.\\

After mass purification, exclusive collection of the $^{229}$Th recoil ions onto a small, nearly point-like collection surface is targeted, in order to allow for installing an efficient vacuum ultra-violet optical system for the final identification of the isomeric fluorescence of $^{229m}$Th. To extract the ions from the QMS and guide them towards the collection surface, a nozzle-like system of 3 electrodes with an opening of 2 mm diameter (7) was developed. The voltages applied here are -2 V, -62 V and -22 V, respectively.

The combined ion extraction and purification efficiency of $\alpha$-recoil ions from the $^{233}$U $\alpha$ decay was determined behind the triodic extraction system (7) by two different detection techniques. In the process of optimization, a micro-channel plate (MCP) detector was used for direct ion counting. After the most efficient operational parameters had been identified, a silicon detector was used to evaluate the efficiency by observing the $\alpha$-decay rate of collected $^{229}$Th ions.

\section{Determination of the $^{233}$U source $\alpha$-recoil ion activity}
Before determining the actual extraction and purification efficiencies of the setup, the effective thorium recoil activity of our $^{233}$U source has to be quantified. For completeness, not only $^{229}$Th, but also all other $\alpha$-recoil ion species of the source will be considered in the following. This allows for a discussion of ion extraction efficiencies as a function of their chemical properties. For this purpose, measurements of the $\alpha$-recoil ion activities were performed. These measurements are then compared with model calculations based on stoppimg powers according to Ziegler, Biersack, Littmark (ZBL, \cite{ZBL}). Two different calculations are performed and compared afterwards. First a fully amorphous source structure is assumed as calculated based on TRIM simulations \cite{Ziegler}. In a second step, a polycrystalline source structure is assumed and simulated with the help of the MDrange program code \cite{mdh}. We find, that the measured high $\alpha$-recoil ion activity of the $^{233}$U source strongly indicates a polycrystalline source structure. Such structures are typically not found in electroplated sources \cite{Hashimoto}, so we attribute this finding to the special production process of the $^{233}$U source via the evaporation technique.

\subsection{The $^{233}$U $\alpha$-recoil ion source}
200 kBq of $^{233}$UF$_4$, evaporated onto an aluminum plate with a diameter of the source area of 20 mm, acts as an $\alpha$-recoil ion emitter. From the density of $^{233}$UF$_4$ (6.6 g/cm$^3$), a source thickness of $(360\pm20)$ nm is inferred. The dominant fraction (84 \%) of $^{233}$U decays via a 4.824 MeV $\alpha$ decay channel, leading to a $^{229}$Th recoil energy of about 84.3 keV.

\begin{figure*}
 \begin{center}
 \resizebox{0.75\textwidth}{!}{%
 \includegraphics{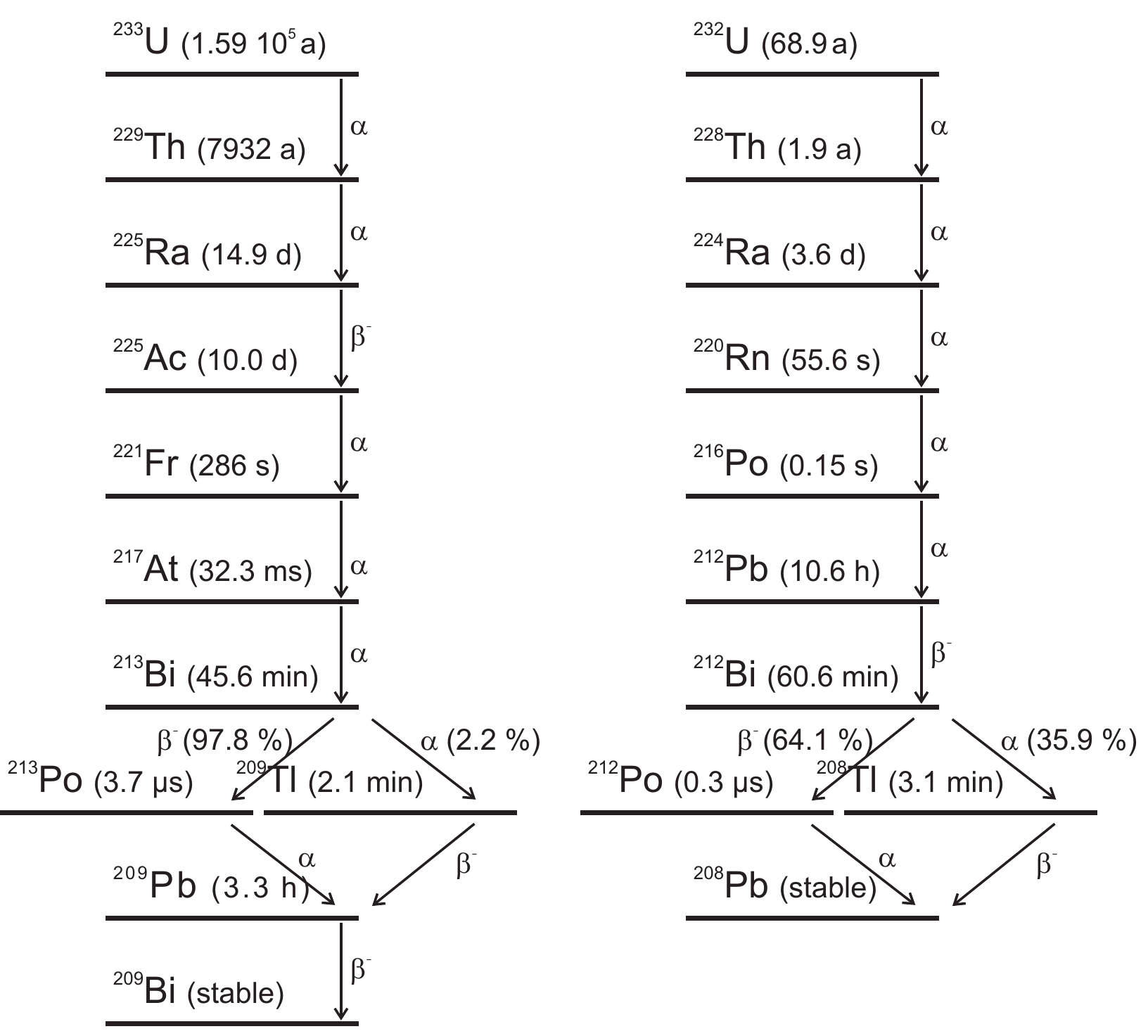}}
 \caption{Decay chains of $^{233}$U and $^{232}$U. $^{232}$U always accompanies 
          $^{233}$U as a small impurity in the source material due to the radiochemical 
          production process.}
 \label{decay_chains}
 \end{center}
\end{figure*}

To fully understand the source, the production process of the $^{233}$U source material is of importance. Typically, highly enriched naturally occuring $^{232}$Th serves as a starting material. This material is then neutron-irradiated in a nuclear reactor. Within this process, $^{233}$U is produced via the reaction chain
\begin{equation}
\resizebox{0.5\textwidth}{!}{$%
^{232}\text{Th}\xrightarrow{(n,\gamma)}\ ^{233}\text{Th}\xrightarrow[21.8\ \text{min}]{(\beta-)}\ ^{233}\text{Pa}\xrightarrow[26.9\ \text{d}]{(\beta-)}\ ^{233}\text{U}. $} 
\end{equation} 
However, there are three side-reaction chains, leading to the production of $^{232}$U and $^{231}$Pa, respectively
\begin{equation}
\resizebox{0.5\textwidth}{!}{$%
\begin{aligned}
^{232}\text{Th}&\xrightarrow{(n,\gamma)}\ ^{233}\text{Th}\xrightarrow[21.8\ \text{min}]{(\beta-)}\ ^{233}\text{Pa}\xrightarrow[26.9\ \text{d}]{(\beta-)}\ ^{233}\text{U}\xrightarrow{(n,2n)}\ ^{232}\text{U}\\
^{232}\text{Th}&\xrightarrow{(n,\gamma)}\ ^{233}\text{Th}\xrightarrow[21.8\ \text{min}]{(\beta-)}\ ^{233}\text{Pa}\xrightarrow{(n,2n)}\ ^{232}\text{Pa}\xrightarrow[1.3\ \text{d}]{(\beta-)}\ ^{232}\text{U}\\
^{232}\text{Th}&\xrightarrow{(n,2n)}\ ^{231}\text{Th}\xrightarrow[25.5\ \text{h}]{(\beta-)}\ ^{231}\text{Pa}.
\end{aligned}$}
\end{equation} 
Further, the $^{232}$Th starting material also contains minor amounts of $^{238}$U, leading to the production of $^{239}$Pu and $^{238}$Pu when neutron irradiated, due to the reactions 
\begin{equation}
\resizebox{0.5\textwidth}{!}{$%
\begin{aligned}
^{238}\text{U}&\xrightarrow{(n,\gamma)}\ ^{239}\text{U}\xrightarrow[23.5\ \text{min}]{(\beta-)}\ ^{239}\text{Np}\xrightarrow[2.36\ \text{d}]{(\beta-)}\ ^{239}\text{Pu}\\
^{238}\text{U}&\xrightarrow{(n,\gamma)}\ ^{239}\text{U}\xrightarrow[23.5\ \text{min}]{(\beta-)}\ ^{239}\text{Np}\xrightarrow[2.36\ \text{d}]{(\beta-)}\ ^{239}\text{Pu}\xrightarrow{(n,2n)}\ ^{238}\text{Pu}\\
^{238}\text{U}&\xrightarrow{(n,\gamma)}\ ^{239}\text{U}\xrightarrow[23.5\ \text{min}]{(\beta-)}\ ^{239}\text{Np}\xrightarrow{(n,2n)}\ ^{238}\text{Np}\xrightarrow[2.1\ \text{d}]{(\beta-)}\ ^{238}\text{Pu}.
\end{aligned}$}
\end{equation} 
While in this context the $^{231}$Pa, $^{238}$Pu and $^{239}$Pu fractions are only of minor importance, the $^{232}$U fraction plays a significant role in the interpretation of the performed measurements. The decay chain products of $^{232}$U are also used for $\alpha$-recoil ion efficiency estimations. The different ion species to be considered for this purpose can be inferred from the decay chains of $^{233}$U and $^{232}$U, respectively, as shown in fig. \ref{decay_chains}.

In order to estimate the source activities of all isotopes contained in the $^{232,233}$U decay chains, the different half-lives involved have to be taken into account. The full $^{232}$U decay chain is inferred to be in equilibrium, while an equilibrium in the $^{233}$U decay chain only occurs from $^{229}$Th downwards. This leads to a significant difference between the activities of $^{233}$U and $^{229}$Th in the source material. Besides the $^{233}$U source activity, the corresponding non-equilibrium factor needs to be known for any estimation of the $\alpha$-recoil ion efficiency. Further, the fractional abundance of $^{232}$U is important for the estimation of the recoil efficiency for the corresponding decay chain. In order to obtain these values, two measurements were performed. First a $\gamma$-ray energy spectrum of the source material was taken in 2007 with the help of a germanium detector (where resolution at the time suffered from neutron damage), second an $\alpha$-ray energy spectrum of the source was taken with a silicon detector (Ametek type BU-017-450-100, thickness 100 $\mu$m, active area 450 mm$^2$) in 2014. Effectively a surface of 22 mm diameter was used for detection.

\begin{figure*}
 \resizebox{1.0\textwidth}{!}{%
 \includegraphics{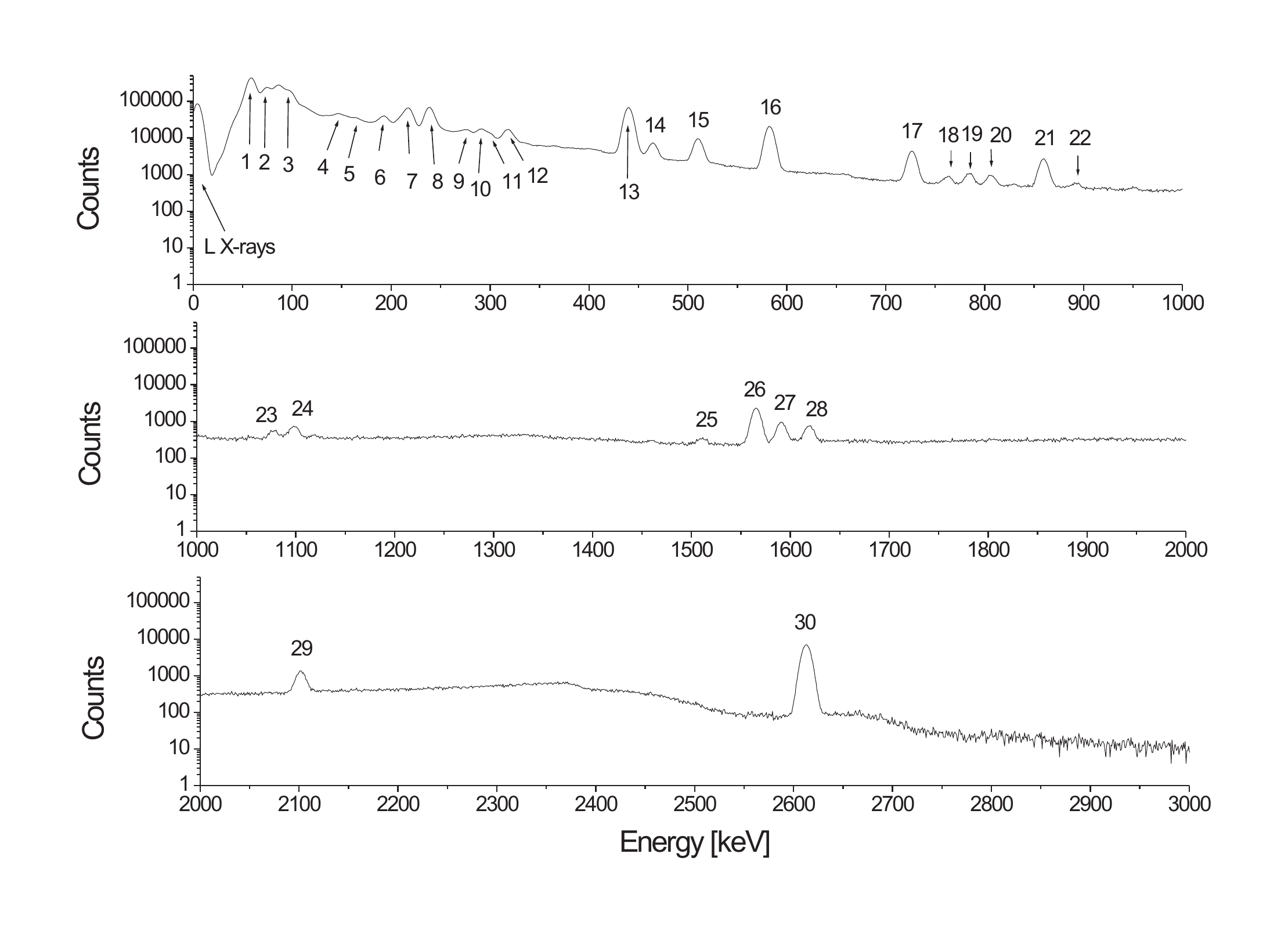}}
 \caption{Full $\gamma$-ray energy spectrum of the $^{233}$U source material. 
          The line assignments are given in table \ref{assignments}.}
 \label{fullspec}
\end{figure*}

The $\gamma$-ray energy spectrum is shown in fig. \ref{fullspec}. It allows for a determination of the $^{229}$Th to $^{228}$Th activity ratio. For this purpose, the integrals over all $\gamma$-ray lines were corrected for the line intensities, branching ratios and detection efficiencies to obtain the relative activities of the nuclei contained in the source. In order to allow for the correction corresponding to the detection efficiencies, the equilibrium of the $^{232}$U decay chain was used. In this way, the activity ratio of the decay chains at the time of the measurement in 2007 was determined to be $R_1$=A$_{^{229}\text{Th}}$/A$_{^{228}\text{Th}}$=$3.3\pm0.2$.

\begin{table*}
\begin{footnotesize}
\begin{center}
\begin{tabular}{|c|c|c|c|c|}
\hline
No & E [keV] & Decay & Transition & Int. [\%]\\
\hline
1 & 42.4 & $^{233}$U $\rightarrow$ $^{229}$Th & $7/2^+\rightarrow 5/2^+$ & 0.0862\\
\hline
2 & 54.7 & $^{233}$U $\rightarrow$ $^{229}$Th & $9/2^+\rightarrow 7/2^+$ & 0.0182\\
\hline
3 & 97.1 & $^{233}$U $\rightarrow$ $^{229}$Th & $9/2^+\rightarrow 5/2^+$ & 0.0203\\
\hline
4 & 146 & $^{233}$U $\rightarrow$ $^{229}$Th &  $5/2^- \rightarrow 5/2^+$ & 0.00657\\
\hline
5 & 164 & $^{233}$U $\rightarrow$ $^{229}$Th & $3/2^- \rightarrow 3/2^+$ & 0.00623\\
\hline
6 & 187 & $^{233}$U $\rightarrow$ $^{229}$Th & $5/2^- \rightarrow 5/2^+$ & 0.0019\\
  & 188 & $^{225}$Ac $\rightarrow$ $^{221}$Fr & $3/2^+\rightarrow 3/2^-$& 0.54\\
  & 193 & $^{229}$Th $\rightarrow$ $^{225}$Ra & $7/2^+\rightarrow 3/2^+$& 4.4 \\
\hline
7 & 208 & $^{233}$U $\rightarrow$ $^{229}$Th & $7/2^-\rightarrow 5/2^+$ & 0.00229\\
  & 210 & $^{229}$Th $\rightarrow$ $^{225}$Ra & $9/2^+\rightarrow 7/2^+$ &2.99 \\
  & 217 & $^{233}$U $\rightarrow$ $^{229}$Th & $5/2^-\rightarrow 5/2^+$ & 0.0032\\
  & 218 & $^{221}$Fr $\rightarrow$ $^{217}$At & $5/2^-\rightarrow 9/2^-$ & 11.6 \\
\hline
8 & 238 & $^{212}$Pb $\rightarrow$ $^{212}$Bi & $0^-\rightarrow 1^-$ & 43.6\\
  & 241 & $^{224}$Ra $\rightarrow$ $^{220}$Rn & $2^+\rightarrow 0^+$ & 4.1\\
  & 245 & $^{233}$U $\rightarrow$ $^{229}$Th & $ 5/2^+\rightarrow 7/2^+$ &  0.00362\\
\hline
9 & 277 & $^{208}$Tl $\rightarrow$ $^{208}$Pb & $4^-\rightarrow 5^-$& 6.3 \\
\hline
10 & 291 & $^{233}$U $\rightarrow$ $^{229}$Th & $5/2^+\rightarrow 5/2^+$ &0.00537\\
\hline 
11 & 300 & $^{212}$Pb $\rightarrow$ $^{212}$Bi & $1^-\rightarrow 2^-$& 3.3\\
\hline
12 & 317 & $^{233}$U $\rightarrow$ $^{229}$Th & $5/2^+\rightarrow 3/2^+$& 0.00776\\
   & 320 & $^{233}$U $\rightarrow$ $^{229}$Th & $5/2^+\rightarrow 5/2^+$& 0.00290\\
   & 323 & $^{213}$Bi $\rightarrow$ $^{213}$Po & $3/2^+\rightarrow 1/2^+$& 0.165 \\
\hline
13 & 440 & $^{213}$Bi $\rightarrow$ $^{213}$Po & $7/2^+\rightarrow 9/2^+$ & 26.1\\
\hline
14 & 465 & $^{209}$Tl $\rightarrow$ $^{209}$Pb & $1/2^+\rightarrow 5/2^+$ & 96.9 \\
\hline
15 & 510 & $^{208}$Tl $\rightarrow$ $^{208}$Pb & $5^-\rightarrow 5^-$& 22.6\\
   & 511 & e$^+$ e$^-$ annihil. & & \\
\hline
 16 & 570 & $^{212}$Po $\rightarrow$ $^{208}$Pb & $5^-\rightarrow 3^-$& 2.0\\
    & 583 & $^{208}$Tl $\rightarrow$ $^{208}$Pb & $5^-\rightarrow 3^-$& 84.5\\
\hline
 17 & 727 & $^{212}$Bi $\rightarrow$ $^{212}$Po & $2^+ \rightarrow 0^+$&  6.7\\
\hline
 18 & 763 & $^{208}$Tl $\rightarrow$ $^{208}$Pb & $5^-\rightarrow 5^-$ & 1.8\\
\hline
 19 & 785 & $^{212}$Bi $\rightarrow$ $^{212}$Po & $2^+\rightarrow 2^+$& 1.1\\
\hline
 20 & 807 & $^{213}$Bi $\rightarrow$ $^{213}$Po & $7/2^+\rightarrow 11/2^+$& 0.29\\
\hline
 21 & 860 & $^{208}$Tl $\rightarrow$ $^{208}$Pb & $4^-\rightarrow 3^-$ & 12.4\\
\hline
 22 & 893 & $^{212}$Bi $\rightarrow$ $^{212}$Po & $1^+ \rightarrow 2^+$ & 0.378\\
\hline
 23 & 1078 & $^{212}$Bi $\rightarrow$ $^{212}$Po & $2^+\rightarrow2^+$ & 0.564\\
\hline
 24 & 1100 & $^{213}$Bi $\rightarrow$ $^{213}$Po & $7/2^+\rightarrow 9/2^+$ &  0.259\\
\hline
 25 & 1512 & $^{212}$Bi $\rightarrow$ $^{212}$Po & $2^+ \rightarrow 0^+$ & 0.29\\
\hline 
 26 & 1567 & $^{209}$Tl $\rightarrow$ $^{209}$Pb & $5/2^+\rightarrow 9/2^+$ &  99.8\\
\hline
 27 & 1592 & double escape & &\\
\hline
 28 & 1620 & $^{212}$Bi $\rightarrow$ $^{212}$Po & $1^+ \rightarrow 0^+$ & 1.47\\
\hline
 29 & 2103 & single escape &&\\
\hline
 30 & 2610 & $^{212}$Po $\rightarrow$ $^{208}$Pb & $3^-\rightarrow 0^+$ & 2.6\\
    & 2614 & $^{208}$Tl $\rightarrow$ $^{208}$Pb & $3^-\rightarrow 0^+$& 99.754\\
\hline

\end{tabular}
\caption{Assignments of the transitions indicated in the $\gamma$-ray energy spectrum 
         shown in fig. \ref{fullspec}. Besides the energies of the transitions, their 
         relative decay intensities are given \cite{gov}.} 
\label{assignments}
\end{center}
\end{footnotesize}
\end{table*}

\begin{figure*}
 \begin{center}
 \resizebox{0.75\textwidth}{!}{%
 \includegraphics[totalheight=9cm]{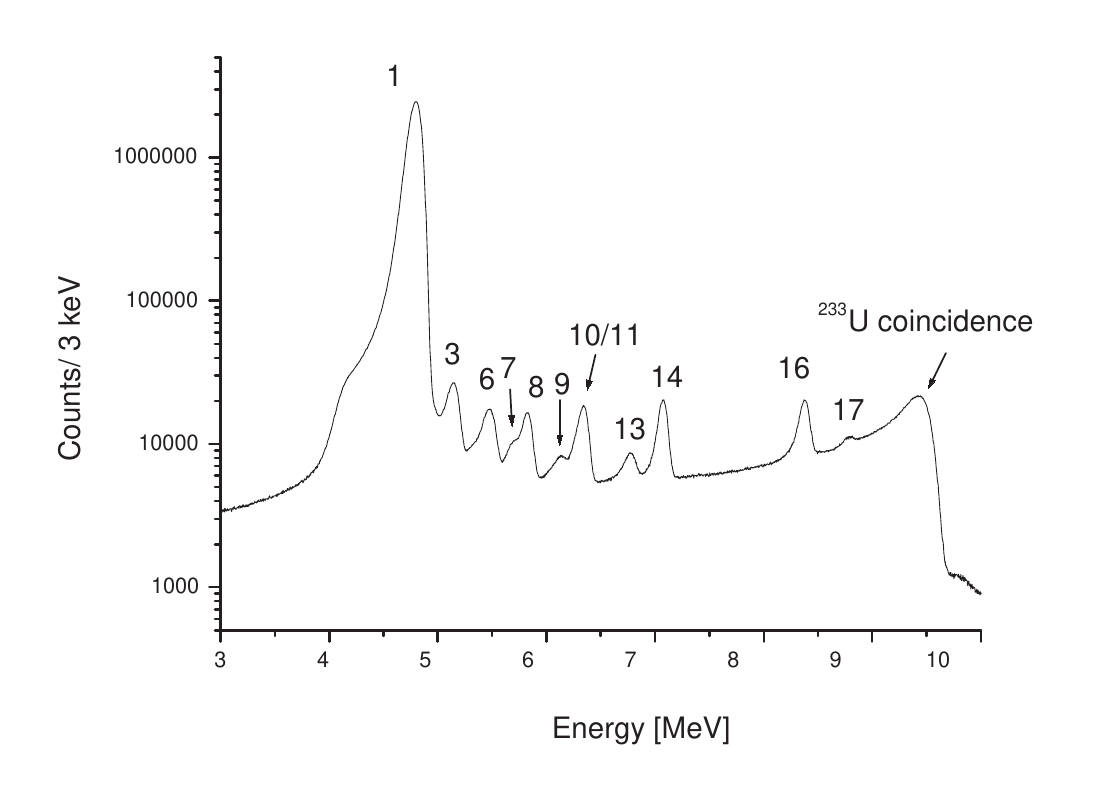}}
 \caption{$\alpha$-energy spectrum as obtained after 1 hour of detection, when the silicon detector was placed in 5 mm distance directly in front of the $^{233}$U source. The line assignments are listed in table \ref{alphaassign}. Besides the decay chains of $^{233}$U and $^{232}$U, also lines of $^{238}$Pu and $^{239}$Pu are visible.}
 \label{alpha_spec}
 \end{center}
\end{figure*}

The direct $\alpha$-energy spectrum, as obtained after 1 hour of measurement at a distance of 5 mm in front of the source, is shown in fig. \ref{alpha_spec}. This spectrum allows for a determination of the time since chemical purification of the source material, as well as of the absolute source activity. For this purpose, the $^{233}$U to $^{229}$Th activity ratio is measured to be $R_2=A_{^{233}\text{U}}/A_{^{229}\text{Th}}=250\pm10$, based on a comparison of lines 1 and 16 in fig. \ref{alpha_spec}. The age of the source material is then calculated to be $t=1/(\lambda_{^{229}\text{Th}}\cdot R_2)=45\pm5$ years and was therefore produced around 1969. The absolute $^{233}$U $\alpha$ activity of the source is obtained to be $(200\pm10)$ kBq, which is slightly below the design value of $228$ kBq. The errors in these measurements are dominated by the knowledge about the distance between the silicon detector and the source, as well as the choices for the integration limits.

\begin{table*}
\begin{footnotesize}
\begin{center}
\begin{tabular}{|c|c|c|c|c|}
\hline
No& Isotope & Half life &  Energy [keV] & Intensity [\%]\\
\hline
1 & $^{233}$U & $159.2\cdot 10^{3}$ y & 4783 & 13.2 \\
 & & & 4824 & 84.3 \\
\hline
2  & $^{229}$Th & 7932 y & 4814 & 9.30 \\
  & & & 4838 & 5.00 \\
  & & & 4845 & 56.20 \\
  & & & 4901 & 10.20 \\
  & & & 4967 & 5.97 \\
  & & & 5053 & 6.60 \\
\hline
 3 & $^{239}$Pu & $2.41\cdot 10^4$ y & 5105 & 11.94 \\
 & & & 5144 & 17.11 \\  
 & & & 5156 & 70.77 \\ 
  \hline
4 & $^{232}$U & 68.9 y& 5263 & 31.55 \\
  & & & 5320 & 68.15 \\  
 \hline
5 & $^{228}$Th & 1.9 y& 5340 & 27.20 \\
  & & & 5423 & 72.20 \\  
\hline
6 & $^{238}$Pu & $87.7$ y & 5456 & 28.98 \\
 & & & 5499 & 70.91 \\ 
\hline
7 & $^{224}$Ra &3.6 d& 5448 & 5.06\\
  & & & 5685 & 94.92\\
\hline
8 & $^{225}$Ac &10.0 d& 5732 & 8.00\\
  & & & 5790 & 8.60 \\
  & & & 5792 & 18.10 \\
  & & & 5830 & 50.70 \\
\hline
9 & $^{212}$Bi &60.55 min& 6051 & 25.13\\
  & & & 6090 & 9.75\\
\hline
10 & $^{220}$Rn &55.6 s& 6288 & 99.89\\
\hline
11 & $^{221}$Fr &286.1 s& 6126 & 15.10\\
  & & & 6341 & 83.40\\
\hline
12 & $^{211}$Bi &2.14 min& 6278 & 16.19\\
  & & & 6623 & 83.54\\
\hline
13 & $^{216}$Po &0.145 s& 6778 & 99.99\\
\hline
14 & $^{217}$At &32.3 ms& 7067 & 99.89\\
\hline
15 & $^{215}$Po &1.78 ms& 7386 & 100\\
\hline
16 & $^{213}$Po &3.72 $\mu$s & 8376 & 100\\
\hline
17 & $^{212}$Po &0.30 $\mu$s& 8785 & 100\\
\hline
\end{tabular}
\end{center}
\caption{List of $\alpha$ decay channels of the isotopes contained in the $^{232,233}$U decay chains, together with their relative intensities taken from \cite{gov}. As far as seen, also lines of the $^{239}$Pu, $^{238}$Pu and $^{231}$Pa decay chains are listed. The lines are listed corresponding to the assignments given in the $\alpha$-energy spectra shown in figs. \ref{alpha_spec}, \ref{siliconrecoil} and \ref{Si_spectrum}. Only lines with more than 5 \% intensity are listed.}
\label{alphaassign}
\end{footnotesize}
\end{table*}

Having all these numbers at hand, the source is quantitatively well under control. The measured activity of about $200$ kBq corresponds to an abolute number of $^{233}$U nuclei of $1.45\cdot 10^{18}$. The $^{229}$Th to $^{228}$Th activity ratio in 2007 $(R_1)$ is used to determine the relative abundance of $^{232}$U in the source material at the time of radio-chemical purification about 45 years ago, making use of the equation 
\begin{equation}
\resizebox{0.4\textwidth}{!}{$%
\begin{aligned}
\frac{N_{^{232}\text{U}}}{N_{^{233}\text{U}}}&= \frac{\lambda_{^{229}\text{Th}}\lambda_{^{233}\text{U}}\left(\lambda_{^{228}\text{Th}}-\lambda_{^{232}\text{U}}\right)t}{R_2\cdot\lambda_{^{228}\text{Th}}\lambda_{^{232}\text{U}}e^{-\lambda_{^{232}\text{U}}t}}\\
&=(6.1\pm0.3)\cdot 10^{-7}.
\end{aligned}$}
\label{uraniumfraction}
\end{equation}
Thus the absolute number of $^{232}$U nuclei at that time was $8.85\cdot 10^{11}$. These parameters are subsequently used as input parameters to model the time evolution of the source activity with the help of the Bateman equation \cite{Bateman}. This equation generally describes the population of isotopic species in a decay chain as populated by one mother nuclide, it reads 
\begin{equation}
\resizebox{0.5\textwidth}{!}{$%
\begin{aligned}
N_k(t)=N_0\left(\prod_{i=1}^{k-1}b_{i}\lambda_i\right)\sum_{j=1}^{k}\frac{\text{e}^{-\lambda_jt}}{\prod_{i=1,i\neq j}^k\left(\lambda_i-\lambda_j\right)}.
\end{aligned}$}
\label{Bateman}
\end{equation}
Here $N_k$ is the number of nuclei of the $k$-th element in the decay chain, $\lambda_k$ is the corresponding decay constant, $b_k$ is the corresponding branching ratio and $N_0$ the starting number of nuclei. The results are shown for 2014 in table \ref{activity}.

\begin{table}
\begin{center}
\begin{footnotesize}

\begin{tabular}{|c|c|c|}
\hline
Isotope & Number & Activity [$s^{-1}$] \\
\hline
$^{233}$U & $1.45 \cdot 10^{18}$ & $2.00 \cdot 10^{5}$  \\
\hline
$^{229}$Th & $2.83 \cdot 10^{14}$  & $7.91 \cdot 10^{2}$  \\
\hline
$^{225}$Ra & $1.46 \cdot 10^{9}$  & $7.90 \cdot 10^{2}$ \\
\hline
$^{225}$Ac & $9.84 \cdot 10^{8}$  & $7.89 \cdot 10^{2}$  \\
\hline
$^{221}$Fr & $3.34 \cdot 10^{5}$  & $7.89 \cdot 10^{2}$  \\
\hline
$^{217}$At & $36.76$              & $7.89 \cdot 10^{2}$   \\
\hline
$^{213}$Bi & $3.12 \cdot 10^{6}$ & $7.89 \cdot 10^{2}$  \\
\hline
$^{213}$Po & $4.67 \cdot 10^{-3}$  & $7.72 \cdot 10^{2}$  \\
\hline
$^{209}$Tl & $3.24 \cdot 10^{3}$ & $17.4$ \\
\hline
$^{209}$Pb & $1.33 \cdot 10^{7}$  & $7.89 \cdot 10^{2}$  \\
\hline
$^{209}$Bi & $5.59 \cdot 10^{11}$ & $0$  \\
\hline
\hline
 $^{232}$U & $5.63 \cdot 10^{11}$ & $1.80 \cdot 10^{2}$  \\
\hline
 $^{228}$Th & $1.61 \cdot 10^{10}$  & $1.85 \cdot 10^{2}$  \\
\hline
 $^{224}$Ra & $8.43 \cdot 10^{7}$  & $1.85 \cdot 10^{2}$  \\
\hline
 $^{220}$Rn & $1.48 \cdot 10^{4}$  & $1.85 \cdot 10^{2}$  \\
\hline
 $^{216}$Po & $40.0$  & $1.85 \cdot 10^{2}$  \\
\hline
 $^{212}$Pb & $1.02 \cdot 10^{7}$ & $1.85 \cdot 10^{2}$  \\
\hline
 $^{212}$Bi & $9.67 \cdot 10^{5}$ & $1.85 \cdot 10^{2}$  \\
\hline
 $^{212}$Po & $5.12 \cdot 10^{-5}$  & $1.19 \cdot 10^{2}$  \\
\hline
 $^{208}$Tl & $1.75 \cdot 10^{4}$ & $66.4$  \\
\hline
 $^{208}$Pb & $3.06 \cdot 10^{11}$ & $0$  \\
\hline
\end{tabular}
\caption{Calculated absolute numbers of nuclei and corresponding activities for all isotopes of the decay chains of $^{233}$U and $^{232}$U, as contained in the $\alpha$-recoil ion source used.}
\label{activity}
\end{footnotesize}
\end{center}
\end{table}

\subsection{Measurement of the $\alpha$-recoil ion activity}
The $\alpha$-recoil ion activity of the $^{233}$U source is significantly smaller than its $\alpha$ activity, as the energies of the recoil ions scale accordingly to their mass as $E_{\text{rec}}=E_\alpha\cdot m_\alpha/m_{\text{rec}}$. Therefore, the travel length of the ions within the source material is short and a significant fraction of the $\alpha$-recoil ions does not leave the source material itself. Measurements were performed to estimate the $\alpha$-recoil ion efficiency of the $^{233}$U source in use, however, the recoil efficiency for $^{229}$Th could not be observed directly. The reason is, that the source material in use has not been radio-chemically purified within about 45 years and the $\alpha$-recoil ions originating from the short-lived isotopes lead to a significant background signal hindering the direct detection of the $^{229}$Th $\alpha$ decay. The main contribution to this background is due to the $^{225}$Ac activity on the silicon detector, successively populated by the $^{225}$Ra decay. This activity is expected to be significantly reduced after about 150 days of decay, which then might allow for a direct measurement of the $^{229}$Th activity on long time scales.

\begin{figure*}
 \begin{center}
 \resizebox{0.75\textwidth}{!}{%
 \includegraphics[totalheight=9cm]{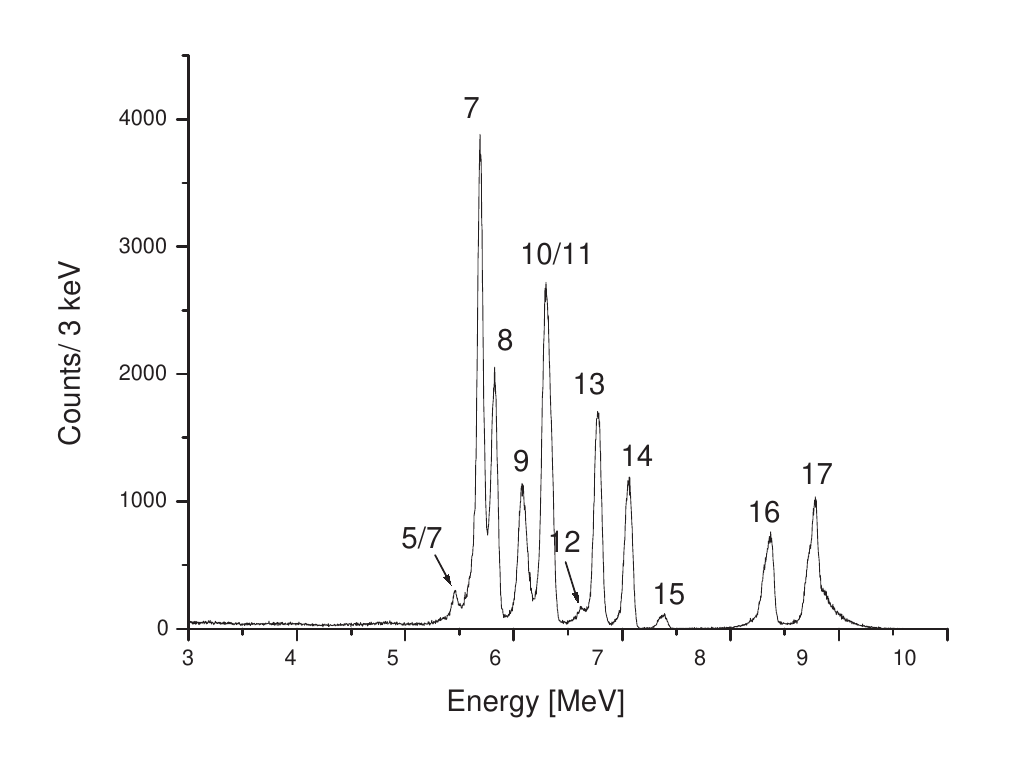}}
 \caption{$\alpha$-energy spectrum as obtained after 5 days of recoil ion implantation, followed by 1 day of decay and 1 further day of detection. The line assignments are listed in table \ref{alphaassign}. $^{229}$Th is not visible with any significance, as the spectrum is heavily dominated by the $^{224}$Ra and $^{225}$Ac decay, respectively. The reason is, that the source material has not been chemically purified to remove the $^{232,233}$U daughters within the past 45 years.}
 \label{siliconrecoil}
 \end{center}
\end{figure*}

Instead, the $\alpha$-recoil ion activities of the $^{233}$U source were determined for some of the shorter lived isotopes in the $^{232,233}$U decay chains, namely $^{225}$Ra, $^{224}$Ra and $^{221}$Fr (please note that $^{225}$Ra, while itself being a $\beta^-$ emitter, is produced via the $\alpha$ decay of $^{229}$Th and therefore the $^{233}$U source does possess a corresponding $\alpha$-recoil activity). The short half-life of $^{221}$Fr did allow for a repetition of the measurement and therefore for some systematic study. The $\alpha$-recoil ions were directly implanted into the silicon detector. For recoil ion implantation, the detector was placed in a central position within 5 mm distance to the $^{233}$U source. The decay of the implanted recoil ions was measured in a sideways position, without a direct line of sight to the source. For this purpose, the silicon detector was mounted movable onto a linear stage. The minimum time required for shifting between both positions was about 60 s, which still allowed for the measurement of the $^{221}$Fr decays. Implantation times of typically 15 min. followed by 15 min. measuring time were chosen. Finally, a long implantation of 5 days was performed, which successively allowed for the determination of the $\alpha$-recoil ion activities for the isotopes of longer half-lives. A corresponding $\alpha$-energy spectrum is shown fig. \ref{siliconrecoil}. This spectrum contains the decays of 1 day of measurement after 5 days of implantation and 1 further day of decay. The decay chains of $^{224}$Ra and $^{225}$Ra are clearly visible, accompanied by the $^{215}$Po and $^{211}$Bi decays, originating from the $^{231}$Pa decay chain. $^{229}$Th, however, cannot be seen with any significance.

When evaluating the $\alpha$-recoil ion implantation measurements, care has to be taken when calculating the number of detected $\alpha$ decays backwards to the number of recoil ions emitted from the source, as the $\alpha$-recoil ions are not emitted into all directions with equal intensity. Instead, the forward direction is strongly favoured, as the recoil ions emitted into this direction have to overcome the smallest amount of source material. The fact that this has indeed a significant effect was shown in a separate measurement, in which the accumulated $\alpha$-recoil ion activity was compared for two different implantation positions of the silicon detector. At first, the silicon detector was positioned 20 mm off axis from the source, at second a central position was chosen. For an assumed isotropic recoil ion distribution, the expected activity ratio between these measurements would be 0.148, based on numerical ray-tracing simulations taking the extension of the source and the detector geometry into account. This value could be confirmed by comparing measurements of the $^{233}$U $\alpha$ decay in the same detector positions, where an activity ratio of 0.142 was obtained. In case of the $^{221}$Fr recoil ions, however, the same ratio reached a value of 0.070, thus strongly indicating an enhancement of the recoil ion activity in the forward direction. Model calculations for the $\alpha$-recoil ion distribution as discussed in sect. \ref{sec:model} were applied to account for this effect.

The absolute numbers of measured $\alpha$-recoil activities are given in table \ref{measuredrecoils}, together with the resulting recoil efficiencies of the $^{233}$U source. It turns out that the measured values are well in agreement and distributed around 5.5 \%. In order to compare these values with theoretical predictions, model calculations were performed and will be discussed in the following section.

\begin{table}
\begin{center}
\begin{footnotesize}
\begin{tabular}{|c|c|c|}
\hline
Isotope &  Rec. act. [$s^{-1}$] & Rec. eff. [\%]\\
\hline
$^{225}$Ra & $42.8\pm4.3$ & $5.4\pm0.5 $  \\
\hline
$^{224}$Ra & $10.2\pm1.0$  &  $5.5\pm0.6 $ \\
\hline
$^{221}$Fr & $45.3\pm4.5$  & $5.7\pm0.6 $  \\
\hline
\end{tabular}
\caption{Measured absolute $\alpha$-recoil activities and recoil efficiencies of the $^{233}$U source.}
\label{measuredrecoils}
\end{footnotesize}
\end{center}
\end{table}

\subsection{Model calculations for $\alpha$-recoil ion efficiencies}
\label{sec:model}
In order to compare the measured $\alpha$-recoil ion efficiencies as given in table \ref{measuredrecoils} with theoretical expectations and to provide realistic estimations for the number of recoil ions of the isotopes, which could not directly be measured, some model calculations were performed. For this purpose, the ZBL stopping power \cite{ZBL} was applied for two different source structures. First a fully amorphous source material was assumed, as simulated with the help of TRIM \cite{Ziegler}, second a more sophisticated model was applied, assuming a polycrystalline UF$_4$ source structure. The latter simulations were performed with the MDrange program code \cite{mdh} for different crystal sizes ranging from 10 to 100 nm. It turned out, that a polycrystalline source structure with $(80\pm10)$ nm crystal size is able to reproduce the measured $\alpha$-recoil efficiencies. With the help of these simulations, values for the stopping length $s$ in the source material, as well as the standard deviations in longitudinal ($\sigma_{\text{lo}}$) and transversal ($\sigma_{\text{tr}}$) directions were obtained. As these values vary with the energy of the recoil ion as well as the atomic number, simulations were performed for each isotope separately. These values were then used to model the probability density of the stopped ions by a 3 dimensional gaussian function. The fraction of ions, which leave the source, is then calculated as a function of the $\alpha$-decay depth $r$ in the source material and the recoil direction described by a polar angle $\theta$ by integrating this gaussian function over the half space, in which no source material exists. In order to account for the dependency on $r$ and $\theta$, the gaussian function is accordingly turned and shifted. The corresponding integral is analytically solved. The result is shown in eq. \ref{transmissionfunction}.
\begin{table*}
\begin{equation}
\resizebox{0.75\textwidth}{!}{$%
\begin{aligned}
T(r,\theta)&=\frac{1}{(2\pi)^{3/2}\sigma_{\text{tr}}^2\sigma_{\text{lo}}}\int\limits_{-\infty}^{\infty}dx\int\limits_{-\infty}^{\infty} dy\int\limits_{\frac{r+x\sin(\theta)}{\cos(\theta)}}^{\infty} e^{-\left(\frac{x^2+y^2}{2\sigma_{\text{tr}}^2}+\frac{(z-s)^2}{2\sigma_{\text{lo}}^2}\right)}\ dz\\
&=\frac{1}{2}\left[1+\text{erf}\left(\frac{s\cos(\theta)-r}{\sqrt{2}\sigma_{\text{tr}}\sin(\theta)}\cdot\left[\left(\frac{\sigma_{\text{lo}}\cos(\theta)}{\sigma_{\text{tr}}\sin(\theta)}\right)^2+1\right]^{-1/2}\right)\right]
\end{aligned}$}
\label{transmissionfunction}
\end{equation}
\end{table*}

\begin{figure}
 \begin{center}
 \resizebox{0.5\textwidth}{!}{%
 \includegraphics[totalheight=5cm]{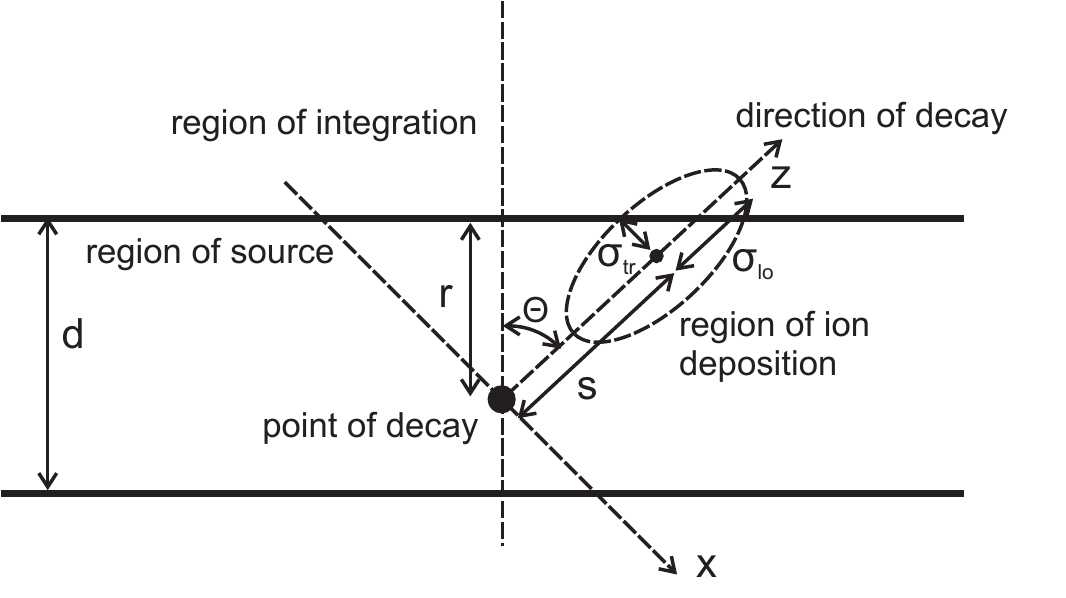}}
 \caption{Visualization of the variables contained in the theoretical model which is 
          applied for an estimation of the $\alpha$-recoil ion activity of the $^{233}$U 
          source.}
 \label{recoilfig}
 \end{center}
\end{figure}

Herein, $erf$ denotes the error function, the meaning of all variables is visualized in fig.~\ref{recoilfig}.
The obtained transmission function of eq. (\ref{transmissionfunction}) was subsequently used to numerically estimate the total fraction of $\alpha$-recoil ions, which are able to leave the source for a given source thickness d via the integral
\begin{equation}
\resizebox{0.5\textwidth}{!}{$%
\begin{aligned}
F=\frac{1}{4\pi d}\int_0^d dr\int_0^{2\pi}d\phi\int_0^{\pi} \sin(\theta)d\theta \ T(r,\theta).
\end{aligned}$}
\label{transmission}
\end{equation} 
In table \ref{simulatedrecoils}, the calulated $F$-values based on the TRIM as well as the MDrange simulations are compared with the measured recoil efficiencies, as obtained in the previous section. Surprisingly, the data based on the TRIM simulations lead to a significant underestimation of the $\alpha$-recoil ion activity of the $^{233}$U source in use.
\begin{table}
\begin{center}
\begin{footnotesize}
\begin{tabular}{|c|c|c|c|}
\hline
Isotope & $F_{\text{meas}}$ & $F_{\text{TRIM}}$ & $F_{\text{MDH}}$\\
\hline
$^{225}$Ra & $5.4 $ \% & $1.9$ \%& $5.5$ \% \\
\hline
$^{224}$Ra & $5.5 $ \% & $2.0$ \%& $5.5$ \% \\
\hline
$^{221}$Fr &  $5.7 $ \% & $2.1$ \%& $5.8$ \% \\
\hline
\end{tabular}
\caption{Measured and simulated ${\alpha}$-recoil efficiencies of the $^{233}$U source. The TRIM simulations underestimate the actual recoil activity by a factor of about 2.8. The measured values are well reproduced by the MDrange simulations, when a polycrystalline source with about $80$ nm crystal size is assumed.} 
\label{simulatedrecoils}
\end{footnotesize}
\end{center}
\end{table}
 In order to investigate this behaviour in more detail, further comparisons of measured and simulated $\alpha$-recoil ion activities were performed. As the described source is the only one at hand, we applied the model calculations to U sources whose $\alpha$-recoil ion efficiencies were obtained by T. Hashimoto {\it et al.} \cite{Hashimoto}. In this investigation eight $^{238}$U sources (UO$_2$ and U$_3$O$_8$ with different $^{235}$U contents) were electrodeposited with different thicknesses onto 20 mm diameter mirror polished stainless steel plates and probed for their $^{234}$Th recoil-ion efficiency. The results of the comparison are listed in table \ref{Hashimotocomp}.
\vspace{-0cm}
\begin{table}
\begin{footnotesize}
\begin{center}
\begin{tabular}{|c|c|c|c|}
\hline
Source &  $d$ [nm] & F$_{\text{meas}}$ & F$_{\text{TRIM}}$    \\
\hline
EU-5 & $11.8$ & $20$ \% &   $30$ \%  \\
\hline
NU-D &  $15.5$ & $17$ \% &   $25$ \%   \\
\hline
EU-6 &  $16.4$ & $13$ \% &   $24$ \%  \\
\hline
EU-2 &  $53.8$ & $5.8$ \% &   $7.8$ \%   \\
\hline
EU-3 &  $78.7$ & $6.3$ \% &    $6.5$ \%    \\
\hline
EU-4 &  $92.7$ & $3.6$ \% &    $4.5$ \%  \\
\hline
NU-C &  $167.9$ & $1.7$ \% &   $2.5$ \%  \\
\hline
NU-B &  $256.4$ & $1.1$ \% &   $1.6$ \%  \\
\hline
\end{tabular}
\caption{Measured and simulated ${\alpha}$-recoil efficiencies for the 8 different $^{238}$U sources investigated in \cite{Hashimoto}. The TRIM simulations systematically overestimate the actual recoil activities. The dominant factor of overestimation is 1.5. The deposited material of all sources is UO$_2$, except for EU-3, for which UO$_2$ was ignited under air to produce U$_3$O$_8$.} 
\label{Hashimotocomp}
\end{center}
\end{footnotesize}
\end{table} 
This comparison reveals, that for all considered sources the calculated $\alpha$-recoil ion efficiencies based on the TRIM simulations systematically overestimate the actually measured efficiencies. This is the opposite of what was found in the investigation of our source. While a reduction of the $\alpha$-recoil intensity might easily be explained, e.g., by some depostion of the solvent used in the process of electrodeposition, an enhancement above the level as calculated based on TRIM simulations cannot as easily be explained. The only explanation we find is based on the conceptionally different production processes of the considered sources. While the sources used in \cite{Hashimoto} were electrodeposited, the $^{233}$U source in use here was evaporated onto the substrate. While the latter process is known to be significantly cleaner, it may also have led to the build up of a polycrystalline surface. This, moreover, would allow for a significantly enhanced recoil activity due to the possibility of channeling. The MDrange simulations take exactly this effect into account and reveal that the assumption of a polycrystalline surface with about 80 nm crystal sizes could indeed explain the measured $\alpha$-recoil ion activity. For this reason, the recoil activities are calculated based on the MDrange simulations. 

For all isotopes of the $^{232,233}$U decay chains, the values for the stopping length $s$ as well as the standard deviations $\sigma_{\text{lo}}$ and $\sigma_{\text{tr}}$ are listed in table \ref{Fvalues} together with the assumed recoil ion energies (only the dominant $\alpha$-decay energies were taken into account) and the fraction of ions $F$, which is able to leave the source material according to eq. (\ref{transmission}) for a given source thickness of $d=$360 nm.
\vspace{-0cm}
\begin{table*}
\begin{footnotesize}
\begin{center}
\begin{tabular}{|c|c|c|c|c|c|c|c|}
\hline
Isotope & E$_\alpha$ [keV] & Int. [\%] &E$_{\text{rec.}}$ [keV] & $s$ [nm] & $\sigma_{\text{lo}}$ [nm] & $\sigma_{\text{tr}}$ [nm] & F  \\
\hline
$^{229}$Th & 4824 & 84.3 & 84.3 & 75.9 & 35.1 & 5.2 & $5.34 \cdot 10^{-2}$  \\
\hline
$^{225}$Ra & 4845 & 56.2 & 86.1 & 77.5 & 35.9 & 5.3 & $5.45 \cdot 10^{-2}$  \\
\hline
$^{221}$Fr & 5830 & 50.7 & 105.5 & 82.4 & 33.5 & 6.4 & $5.78 \cdot 10^{-2}$  \\
\hline
$^{217}$At & 6341 & 83.4 & 116.9 & 84.6 & 34.7 & 6.2 & $5.93 \cdot 10^{-2}$  \\
\hline
$^{213}$Bi & 7067 & 99.9 & 132.7 & 92.8 & 37.2 & 7.4 & $6.51 \cdot 10^{-2}$  \\
\hline
$^{209}$Tl & 5869 & 1.9 & 112.3 & 85.8 & 35.7 & 6.5 & $6.02 \cdot 10^{-2}$  \\
\hline
$^{209}$Pb & 8376 & 100 & 160.3 & 97.1 & 37.6 & 8.4 & $6.82 \cdot 10^{-2}$  \\
\hline\hline
$^{228}$Th & 5320 & 68.2 & 93.3 & 78.4 & 33.6 & 5.4 & $5.50 \cdot 10^{-2}$  \\
\hline
$^{224}$Ra & 5423 & 72.2 & 96.8 & 78.6 & 34.4 & 5.6 & $5.52 \cdot 10^{-2}$  \\
\hline
$^{220}$Rn & 5685 & 94.9 & 103.4 & 81.0 & 34.1 & 6.3 & $5.69 \cdot 10^{-2}$  \\
\hline
$^{216}$Po & 6288 & 99.9 & 116.4 & 85.3 & 34.7 & 6.3 & $5.98 \cdot 10^{-2}$  \\
\hline
$^{212}$Pb & 6778 & 100 & 127.9 & 87.8 & 35.4 & 7.0 & $6.16 \cdot 10^{-2}$  \\
\hline
$^{208}$Tl & 6051 & 25.1 & 116.4 & 87.4 & 34.6 & 6.5 & $6.13 \cdot 10^{-2}$  \\
\hline
$^{208}$Pb & 8785 & 100 & 168.9 & 99.3 & 39.2 & 8.9 & $6.98 \cdot 10^{-2}$  \\
\hline
\end{tabular}
\caption{List of $\alpha$-recoil ions with corresponding dominant $\alpha$-decay energies, intensities and $\alpha$-recoil ion energies, together with stopping lengths $s$ and standard deviations $\sigma_{\text{lo}}$ and  $\sigma_{\text{tr}}$ as obtained by MDrange simulations \cite{mdh}. Note, that the $\alpha$-decay energies as listed in the table correspond to the mother nuclide (e.g. $^{233}$U in case of $^{229}$Th). The fractions of $\alpha$-recoil ions, 
         which are able to leave the $^{233}$U source $F$, were calculated in accordance 
         with eq. (\ref{transmission}).} 
\label{Fvalues}
\end{center}
\end{footnotesize}
\end{table*}

The estimated recoil efficiency F for $^{229}$Th amounts to $5.34\cdot 10^{-2}$, leading to an absolute number of $(10.7\pm 1.1)\cdot 10^{3}$ recoil ions per second entering the active volume of the buffer-gas stopping cell. Here, the error was estimated to be 10\%. An overview of the obtained values of activity, $\alpha$-recoil efficiency $F$ and absolute $\alpha$-recoil activity $\epsilon$ of the $^{233}$U source is given for all isotopes in table \ref{activity2}. 
\begin{table*}
\begin{center}
\begin{footnotesize}

\begin{tabular}{|c|c|c|c|}
\hline
Isotope & Activity [$s^{-1}$] & Recoil efficiency F & Recoil activity $\epsilon$ [$s^{-1}$]\\
\hline
$^{233}$U & $2.00 \cdot 10^{5}$ & $0$ & $0$ \\
\hline
$^{229}$Th & $7.91 \cdot 10^{2}$  & $5.34 \cdot 10^{-2}$ & $(10.7\pm1.1)\cdot 10^3$ \\
\hline
$^{225}$Ra & $7.90 \cdot 10^{2}$  & $5.45 \cdot 10^{-2}$ & $43.1\pm4.3$ \\
\hline
$^{225}$Ac & $7.89 \cdot 10^{2}$  & $0$ & (produced by $\beta$ decay) \\
\hline
$^{221}$Fr & $7.89 \cdot 10^{2}$ & $5.78 \cdot 10^{-2}$ & $45.6\pm4.6$ \\
\hline
$^{217}$At & $7.89 \cdot 10^{2}$ & $5.93 \cdot 10^{-2}$  & $46.8\pm4.7$ \\
\hline
$^{213}$Bi & $7.89 \cdot 10^{2}$ & $6.51 \cdot 10^{-2}$ &$51.4\pm5.1$ \\
\hline
$^{213}$Po & $7.72 \cdot 10^{2}$  & $0$ & (produced by $\beta$ decay) \\
\hline
$^{209}$Tl & $17.4$ & $6.02 \cdot 10^{-2}$ & $1.0\pm0.1$ \\
\hline
$^{209}$Pb & $7.89 \cdot 10^{2}$  & $6.82 \cdot 10^{-2}$ &  $52.7\pm5.3$ \\
\hline
$^{209}$Bi & $0$ & $0$ & (produced by $\beta$ decay) \\
\hline
\hline
 $^{232}$U & $1.80 \cdot 10^{2}$ & $0$ & $0$ \\
\hline
 $^{228}$Th & $1.85 \cdot 10^{2}$  & $5.50 \cdot 10^{-2}$ & $9.9\pm1.0$ \\
\hline
 $^{224}$Ra & $1.85 \cdot 10^{2}$  & $5.52 \cdot 10^{-2}$ & $10.2\pm1.0$ \\
\hline
 $^{220}$Rn & $1.85 \cdot 10^{2}$  & $5.69 \cdot 10^{-2}$ & $10.5\pm1.1$ \\
\hline
 $^{216}$Po & $1.85 \cdot 10^{2}$ & $5.98 \cdot 10^{-2}$ & $11.1\pm1.1$\\
\hline
 $^{212}$Pb & $1.85 \cdot 10^{2}$ & $6.16 \cdot 10^{-2}$ & $11.4\pm1.1$ \\
\hline
 $^{212}$Bi & $1.85 \cdot 10^{2}$ & $0$ & (produced by $\beta$ decay) \\
\hline
 $^{212}$Po & $1.19 \cdot 10^{2}$  & $0$ & (produced by $\beta$ decay) \\
\hline
 $^{208}$Tl & $66.4$ & $6.13 \cdot 10^{-2}$ & $4.1\pm0.4$ \\
\hline
 $^{208}$Pb & $0$ & $6.98 \cdot 10^{-2}$ & $8.3\pm0.8$ \\
\hline
\end{tabular}
\caption{Calculated $\alpha$ activities as well as $\alpha$-recoil efficiencies 
         and absolute $\alpha$-recoil activities for all isotopes of the decay chains of 
         $^{233}$U and $^{232}$U, as contained in the $\alpha$-recoil ion source used.}
\label{activity2}
\end{footnotesize}
\end{center}
\end{table*}

It should be pointed out that for our source there is no significant transmission above a source thickness of 150 nm. Typically, this value is even a factor of 3 smaller. As the source in use here is significantly thicker, the error of the recoil efficiency is not influenced by the error of the source activity. Obviously, a thinner $^{233}$U source would be advantageous from the point of view of the recoil efficiency. Indeed, it is planned to replace the existing source with another one of 260 kBq activity, electrodeposited onto a larger surface area of 90 mm diameter (limited by the inner funnel diameter of 115 mm). For this source a thickness of 16 nm could be achieved for U$_3$O$_8$. However, as the technique of source evaporation is not available anymore (the radioactive target laboratory with its evaporation facilities was dismantled in the mean time and no other facility allowing for evaporating nuclear fuel-type target materials is known to us), the envisaged production process of electrodeposition will lead to a significant reduction compared to the maximum $\alpha$-recoil activity that could theoretically be reached. Based on the assumption of an amorphous source, the maximum achievable recoil efficiency (based on TRIM simulations for U$_3$O$_8$) is $2.8\cdot 10^{-1}$, thus leading to an absolute recoil activity of 72800 s$^{-1}$, which is a factor of 6.8 more intense than the present source.\\

\subsection{Implanted $\alpha$-recoil ion activity}
As it turns out, in addition to the $\alpha$-recoil ions directly emitted from the $^{233}$U source, also the $\alpha$-recoil ions as emitted from the source's surrounding due to recoil ion implantation play a non-negligible role. While the buffer-gas stopping cell is held under vacuum, the $\alpha$-recoil ions are continuously implanted into the surrounding metal surfaces. Due to the implantations, these surfaces themselves become $\alpha$-recoil ion emitters. In the following, the significance of this effect will be estimated.

$\alpha$-recoil ions of all isotopic species as contained in the $^{232,233}$U decay chains and emitted by the source due to $\alpha$ decay are implanted into the surrounding metal surfaces in parallel. If we denote the implantation rate for the $k$-th isotope of the decay chain as $\epsilon_k$, the system of differential equations to be solved in order to estimate the implanted activity reads
\begin{equation}
\resizebox{0.5\textwidth}{!}{$%
\begin{aligned}
\dot{N}_1(t)&=-\lambda_1 N_1(t)+\epsilon_1\\
\dot{N}_2(t)&=b_1\lambda_1 N_1(t)-\lambda_2 N_2(t)+\epsilon_2\\
&\vdots\\
\dot{N}_k(t)&=b_{k-1}\lambda_{k-1} N_{k-1}(t)-\lambda_k N_k(t)+\epsilon_k.
\end{aligned}$}
\label{diffsys}
\end{equation}
The solution to this system of differential equations is the Bateman equation with source terms $\epsilon_k$,
which reads for the case $N_k(0)=0$ for all $k$
\begin{equation}
\resizebox{0.5\textwidth}{!}{$%
\begin{aligned}
N_k^{\text{acc}}(t)=\sum_{i=1}^{k}\left(\prod_{l=i}^{k-1}b_l\lambda_l\right)\sum_{j=i}^{k}\frac{\epsilon_i\left(1-e^{-\lambda_jt}\right)}{\lambda_j\prod_{l=i,l\neq j}^k(\lambda_l-\lambda_j)}.
\end{aligned}$}
\label{solution1}
\end{equation}
As soon as buffer gas enters the system, the accumulation of $\alpha$-recoil ions into the surrounding surface stops. At that point the usual isotopic decay takes over, as described by the Bateman equation (\ref{Bateman}), but this time not only one isotope populates the decay chain. Instead, there are non-zero starting values $N_k(0)\neq0$ for all isotopes of the chain. This is taken into account by successive summation over the Bateman equations, describing the population caused by individual starting isotopes. The corresponding solution formula reads
\begin{equation}
\resizebox{0.5\textwidth}{!}{$%
\begin{aligned}
N_k^{\text{dec}}(t)=\sum_{i=1}^k\left(\prod_{l=i}^{k-1}b_l\lambda_l\right)\sum_{j=i}^{k}\frac{N_i(0)e^{-\lambda_jt}}{\prod_{l=i,l\neq j}^k(\lambda_l-\lambda_j)}.
\end{aligned}$}
\label{solution2}
\end{equation}
The complete solution of the system of differential equations (\ref{diffsys}) is given as the sum of the solution formulas (\ref{solution1}) and (\ref{solution2})\footnote{Please note, that a typo was made in \cite{Magill}, where the first product takes values from $1$ to $k-1$ instead of $i$ to $k-1$, which would be correct.}
\begin{equation}
\resizebox{0.5\textwidth}{!}{$%
\begin{aligned}
N_k(t)=&\sum_{i=1}^k\left(\prod_{l=i}^{k-1}b_l\lambda_l\right)\sum_{j=i}^k\Biggr\lbrack\frac{N_i(0)e^{-\lambda_jt}}{\prod_{l=i,l\neq j}(\lambda_l-\lambda_j)}\\
&\qquad+\frac{\epsilon_i\left(1-e^{-\lambda_jt}\right)}{\lambda_j\prod_{l=i,l\neq j}^k(\lambda_l-\lambda_j)}\Biggr\rbrack.
\end{aligned}$}
\end{equation}
These equations are used for the calculation of the implanted source activity. For $\epsilon_k$, the recoil activities as listed in table \ref{activity2} are used. It is assumed that due to the short implantation depths, about $1/4$ of the $\alpha$-recoil ions is able to leave the metal surface. These ions are again implanted into the surrounding metal surfaces, so that they still contribute to the population of the next decay level. The calculation reveals that after about 150 days the surface activity has reached an equilibrium state dominated by the $^{225}$Ra and $^{225}$Ac decays. The numbers of accumulated isotopes $N_{0}$, as well as activities $A_{0}$ and recoil activities $\epsilon_{0}$ as obtained after 150 days of continuous accumulation, are listed in table \ref{surfacerecoils}. As soon as buffer gas enters the system, the $\alpha$-recoil ion implantation stops and the surface activity starts to decay in accordance with eq. (\ref{solution2}). This decay leads to a time dependence of the implanted recoil activity. Isotopes with short half lives are fast decaying, until they reach an equilibrium in accordance with the longer half lives of $^{225}$Ra and $^{225}$Ac. The activities $A_{\text{surf}}$ and recoil activities $\epsilon_{\text{surf}}$ after 1 hour of free decay are also listed in table \ref{surfacerecoils}. 1 hour is chosen as being a typical time at which measurements were started, and also a quasi-equilibrium state is reached for the recoil rates of all isotopes which were actually measured. This time, the branching ratios $b_k$ were used to account for the fact that for all $\alpha$ decays 1/4 of the recoil ions leave the surface and therefore do not populate the next decay level. Finally, the surface recoil activities $\epsilon_{\text{surf}}$ are added to the recoil activities $\epsilon$ of the $^{233}$U source as listed in table \ref{activity2} to give the total expected $\alpha$-recoil ion rate $\epsilon_{\text{total}}$. This analytic discussion reveals that the surface activities are non-negligible.
\begin{table*}
\begin{footnotesize}
\begin{center}
\begin{tabular}{|c|c|c|c||c|c||c|}
\hline
Isotope & $N_0$ & $A_0$ [$s^{-1}$] & $\epsilon_0$ [$s^{-1}$] &   $A_{\text{surf}}$ [$s^{-1}$]& $\epsilon_{\text{surf}}$ [$s^{-1}$] & $\epsilon_{\text{total}}$ [$s^{-1}$]\\
\hline
$^{233}$U & $0$ & $0$ & $0$ &   $0$ & $0$ & $0$\\
\hline
$^{229}$Th & $1.39\cdot10^{11}$  & $0.39$ & $0$ &   $0.39$ &$0$& $(10.7\pm1.1)\cdot 10^3$\\
\hline
$^{225}$Ra & $8.00\cdot10^{7}$  & $43.4$ & $0.098$ &   $43.3$ & $0.098$& $43.1\pm4.3$\\
\hline
$^{225}$Ac & $5.40\cdot10^{7}$  & $43.3$ & $0$ & $43.3$ & $0$ & $0$\\
\hline
$^{221}$Fr & $3.77\cdot10^{4}$  & $88.9$ & $10.8$ & $32.5$ & $10.8$ & $56.4\pm5.6$\\
\hline
$^{217}$At & $6.32$             & $136$ & $22.2$  & $24.4$ & $8.1$ & $54.9\pm5.5$\\
\hline
$^{213}$Bi & $7.39\cdot10^{5}$ & $187$ & $34.0$ & $87.7$ & $6.1$ & $57.5\pm5.8$\\
\hline
$^{213}$Po & $1.11\cdot10^{-3}$  & $183$ & $0$ & $85.7$ & $0$ & $0$\\
\hline
$^{209}$Tl & $7.73\cdot10^{2}$ & $4.13$ & $1.03$ & $1.50$ & $0.48$ & $1.5\pm0.2$\\
\hline
$^{209}$Pb & $4.07\cdot10^{6}$  & $241$ & $46.0$ & $213$ & $21.4$ & $74.1\pm7.4$\\
\hline
$^{209}$Bi & $2.98\cdot10^9$ & $0$ & $0$ & $0$& $0$ & $0$\\
\hline
\hline
 $^{232}$U & $0$ & $0$ & $0$ & $0$ & $0$ & $0$\\
\hline
 $^{228}$Th & $1.19\cdot10^{8}$  & $1.37$& $0$ &  $1.37$& $0$ & $9.9\pm1.0$\\
\hline
 $^{224}$Ra & $5.26\cdot10^{6}$  & $11.5$& $0.34$ & $11.4$& $0.34$ & $10.5\pm1.1$\\
\hline
 $^{220}$Rn & $1.76\cdot10^{3}$  & $22.0$& $2.88$ & $8.58$ & $2.85$ & $13.4\pm1.3$\\
\hline
$^{216}$Po & $7.17$              & $33.1$& $5.50$ & $6.43$ & $2.15$ & $13.3\pm1.3$\\
\hline
 $^{212}$Pb & $2.46\cdot10^{6}$ & $44.5$& $8.28$ & $42.0$ & $1.61$ & $13.0\pm1.3$\\
\hline
 $^{212}$Bi & $2.33\cdot10^{5}$ & $44.5$& $0$ & $43.8$ & $0$ & $0$\\
\hline
 $^{212}$Po &  $1.24\cdot10^{-5}$ & $28.5$ & $0$ & $28.4$ & $0$ & $0$\\
\hline
 $^{208}$Tl & $4.62\cdot10^{3}$ & $17.4$& $3.99$ & $11.8$ & $3.93$ & $8.0\pm0.8$\\
\hline
$^{208}$Pb & $7.22\cdot10^{8}$ & $0$& $7.8$ & $0$ & $7.1$ & $15.4\pm1.5$\\
\hline
\end{tabular}
\caption{Calculated numbers ($N_0$) of isotopes as contained in the metal surfaces of the surrounding of the $^{233}$U source due to $\alpha$-recoil ion implantation. The values are given together with the corresponding activities ($A_0$) and estimated $\alpha$-recoil ion rates ($\epsilon_0$) after 150 days of continuous implantation, when an equilibrium has been reached. The values $A_{\text{surf}}$ and $\epsilon_{\text{surf}}$ give the calcuated activities and $\alpha$-recoil activities of the surface 1 hour after the accumulation was stopped due to buffer gas inlet. The values $\epsilon_{\text{surf}}$ were added to the calculated $^{233}$U source $\alpha$-recoil ion activites $\epsilon$ as listed in table \ref{activity2}, in order to yield the total estimated $\alpha$-recoil ion activity $\epsilon_{\text{total}}$ of the system .}
\label{surfacerecoils}
\end{center}
\end{footnotesize}
\end{table*}

The results will be compared with extraction measurements in sect. \ref{sec:2}, in order to quantify the extraction efficiency. All measurements were taken with the detector system placed behind the extraction electrodes (7) in fig. 1. In this way, a combined extraction and mass-purification efficiency was determined.

\subsection{The special case of $^{220}$Rn}
In the above considerations the special case of $^{220}$Rn as a noble gas was neglected. $^{220}$Rn, as contained in the decay chain of $^{232}$U, could potentially outgas from the bulk material of the source and therefore significantly enhance the extracted amount of $^{216}$Po, while at the same time reduce the extracted amounts of all remaining daughter isotopes.

 As any $^{220}$Rn, which outgasses from the source, is expected to be neutral, it will not be extracted from the stopping cell by the electrode system. Instead it will be floating around in the gas cell, until it is randomly extracted with the He buffer gas. From Laval-nozzle theory \cite{Laval}, the gas flow through the nozzle $\Phi_{\text{He}}$ is known to be $1.70\cdot 10^{20}$ He atoms per second at 40 mbar cell pressure. Comparing this to the absolute number of He atoms in the stopping cell $N_{\text{He}}=1.60\cdot 10^{22}$ leads to the fraction of extracted atoms per second $\lambda_\text{extr}=\Phi_{\text{He}}/N_{\text{He}}=0.0106\ s^{-1}$. This allows one to calculate the expected number of Rn atoms as contained in the He buffer gas in equilibrium $N_{\text{Rn}}$. The corresponding differential equation to be solved is
\begin{equation}
\frac{dN_{\text{Rn}}(t)}{dt}=-\left(\lambda_\text{dec}+\lambda_\text{extr}\right)N_{\text{Rn}}(t)+\epsilon_\text{sour}.
\end{equation}
Herein $\lambda_\text{dec}$ and $\lambda_\text{extr}$ take account for the loss of Rn atoms due to decay and extraction, respectively, while $\epsilon_\text{sour}$ is the term of constant Rn production from the source (185 s$^{-1}$ if all Rn atoms were able to leave the source material). Solving this equation and taking the limit $t\to\infty$ leads to the number of Rn atoms in equilibrium to be
\begin{equation}
N_\text{Rn,equ}=\frac{\epsilon_\text{sour}}{\lambda_\text{dec}+\lambda_\text{extr}}=8.0\cdot10^{3}.
\end{equation}
The predicted Rn activity in the He gas in case of degassing would therefore be $A_\text{Rn}=\lambda_\text{dec}\cdot N_\text{Rn,equ}=100\ s^{-1}$. The $^{216}$Po isotopes produced in these decays are expected to be stopped in the He buffer gas and, as opposed to the Rn, they will be charged and therefore efficiently extracted by the electrode system. An extraction rate of up to $100\ s^{-1}$ for $^{216}$Po can be expected, which is a factor of 7.5 above the value of $13.3\ s^{-1}$ as predicted without considering the degassing of $^{220}$Rn from the source. Further, as practically all $^{216}$Po should be extracted, the extraction rates of the daughter products are expected to be significantly reduced.

Experimentally, however, no enhanced extraction rate of $^{216}$Po is detected and instead the system behaves in a way as if no outgassing of $^{220}$Rn from the source occurs. To explain this behaviour, ultimately the diffusion of Rn in UF$_4$ needs to be discussed. However, there is no data available in literature which would allow for a proper discussion of the diffusion rate. Instead, at this point an upper limit of the expected Rn diffusion is estimated by comparison with the diffusion of Xe in UO$_2$. The Xe diffusion coefficient in UO$_2$ has been well measured to be below $10^{-17}$ cm$^2$s$^{-1}$ at a temperature below 700 K \cite{Matzke}. As the atomic radius of Rn is larger than that of Xe and the temperatures in our setup are significantly lower, this value is taken as an upper limit for the diffusion coefficient of Rn in UF$_4$. The diffusion flux $J$ is calculated for the known diffusion coefficient $D$ from the concentration gradient $dC/dx$ as $J=-D\cdot dC/dx$ \cite{Fick}. The maximum gradient is obtained, if no diffusion occurs. In this case $1.5\cdot10^4$ Rn atoms are contained in the source material of a volume of $1.1\cdot 10^{-4}$ cm$^{3}$. Assuming for simplicity, that a linear density decrease starts to occur at 50 nm distance to the surface, one obtains $-dC/dx\approx 2.7\cdot 10^{13}$ cm$^{-4}$. Thus, the particle flux through the surface of the source caused by diffusion is estimated to be below $2.7\cdot 10^{-4}$ Rn atoms per second and therefore the release of Rn from the source is significantly dominated by the recoil process. No corrections to the results as given in table \ref{surfacerecoils} have to be applied.\\

\section{Determination of the extraction efficiencies}
\label{sec:2}
It is obvious that a large total extraction and purification efficiency for $^{229}$Th from the buffer-gas stopping cell is of major importance for the success of the experimental approach to directly measure the isomeric to ground-state decay of $^{229m}$Th. So far, only estimates could be given \cite{Wense}, based on our own experience with the stopping and extraction of elements other than thorium and based on the findings by the IGISOL group in Jyv\"askyl\"a \cite{Sonnenschein}. Here we report on a measurement of the combined extraction and mass-purification efficiency for $^{229}$Th$^{3+}$ ions from the MLL-IonCatcher buffer-gas stopping cell. Also all other $\alpha$-recoil ion species in the $^{232,233}$U decay chains are considered.

\subsection{MCP detector measurements}
In order to allow for a direct quantification of the extraction efficiency of $^{229}$Th ions, an MCP detector (Hamamatsu, type F2223, 27 mm effective diameter) was placed on the ion extraction axis behind the extraction electrodes (7) in Fig. 1 and was operated in a single-particle counting mode. The applied voltage was -2 kV, such that the positively charged ions were attracted by the MCP surface. A mass spectrum from 10 to 290 u was acquired. The QMS was operated at its resonance frequency of 925 kHz in the mass region from 10 to 200 u (the operational RF amplitude was varied from 78.71 V$_{pp}$ to 1574 V$_{\text{pp}}$). In order to allow also for a scan of the higher mass region, the resonance frequency was lowered to 825 kHz at the expense of mass-resolving power by using a larger inductivity. As the RF voltage amplitude, required for scanning a given mass over charge ratio, depends on the square of the frequency applied to the system, this enabled a mass scan up to 290 u (1816 V$_{\text{pp}}$). The operational parameters of the QMS used for $^{229}$Th ion extraction in different charge states are listed in table \ref{opera}.\\
\vspace{-0cm}
\begin{table}
\begin{footnotesize}
\begin{center}
\begin{tabular}{|c|c|c|c|}
\hline
q & m/q & $\Delta$m/q & m/$\Delta$m \\
\hline
1+ & 229 u/e & 1.5 u/e & $\sim$150 \\
\hline
2+ & 114.5 u/e & 0.8 u/e & $\sim$150 \\
\hline
3+ & 76.3 u/e & 0.5 u/e & $\sim$150 \\
\hline
\hline
q & frequency & RF ampl. & DC volt.\\
\hline
1+ & 825 kHz & 1434 V$_{\text{pp}}$ & 119.7 V\\
\hline
2+ & 925 kHz & 901.2 V$_{\text{pp}}$& 75.23 V\\
\hline
3+ & 925 kHz & 600.5 V$_{\text{pp}}$ & 50.15 V\\
\hline
\end{tabular}
\caption{Operational parameters of the QMS as used for the $^{229}$Th ion extraction 
         listed for the different charge states.} 
\label{opera}
\end{center}
\end{footnotesize}
\end{table}

\begin{figure*}
 \resizebox{1.0\textwidth}{!}{%
 \includegraphics{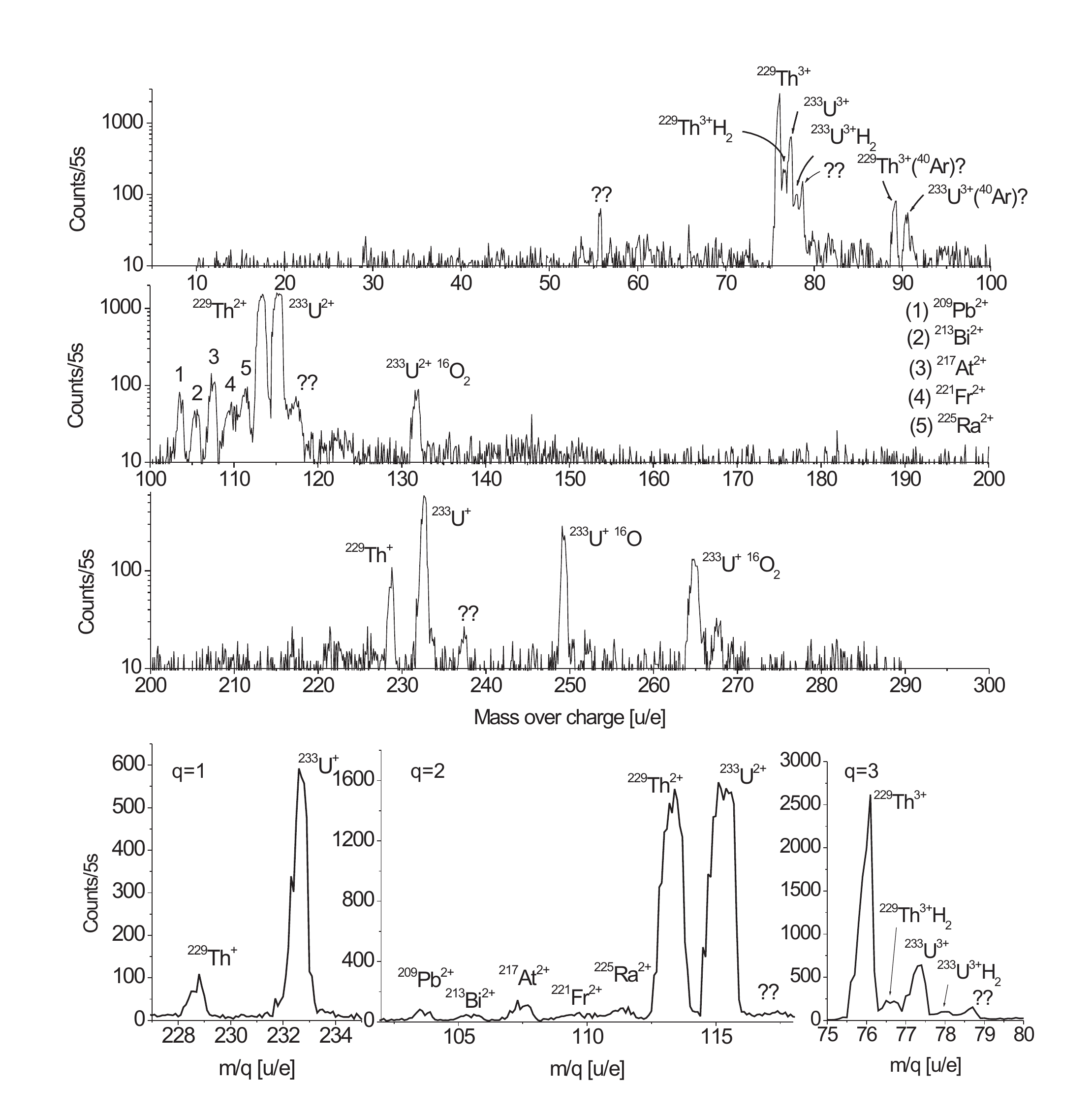}}
 \caption{Mass scan of the extracted recoil ions from 10 to 290 u. The 1+, 2+ and 3+ charged species are clearly visible and shown as zoom inserts below the scan. Minor shifts of the lines occur due to temperature changes during the mass scan.}
 \label{mass_scan}
\end{figure*}

The obtained mass spectra are shown in fig. \ref{mass_scan}. Most of the peaks in the scan could be unambiguously identified. Unidentified peaks or doubtful assignments are indicated by question marks. 

\begin{figure}
 \resizebox{0.5\textwidth}{!}{%
 \includegraphics{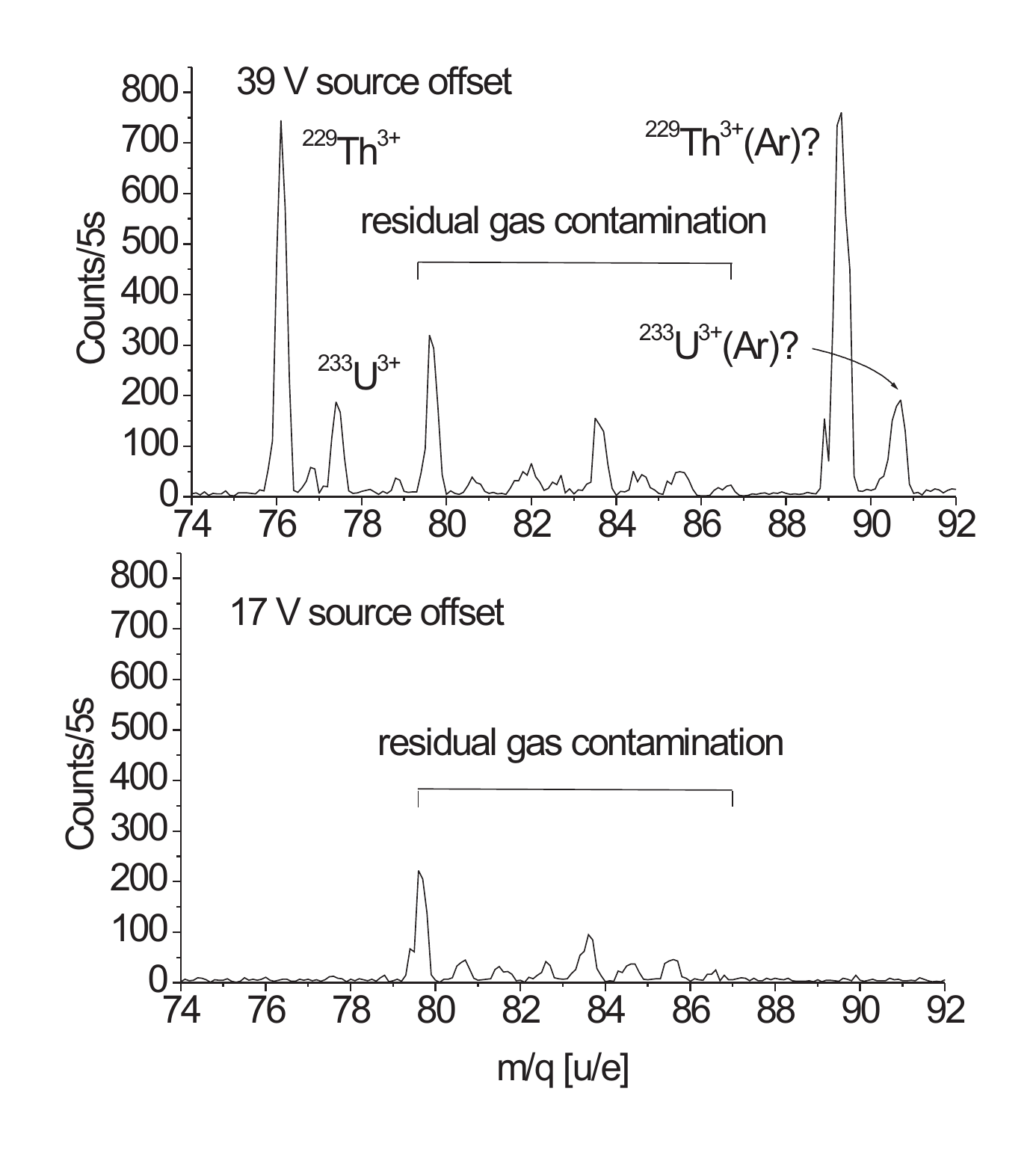}
 }
 \caption{Mass scans for different source-voltage offsets. As can clearly be seen, the extraction of $\alpha$-recoil ions originating from the source is hindered by applying 17 V source offset. The scans were performed shortly after a He bottle change, when still some amount of residual gas contamination was evident.}
 \label{sourceoffsetlabel}
\end{figure}

All assigned peaks could be shown to originate from the $^{233}$U source. This was done by performing a reference scan, for which the source offset was set to 17 V (instead of the 39 V usually used for extraction) and therefore significantly lower than the voltage applied to the funnel entrance of 35 V. In this way, all ions originating from the source were repelled, while other ions were still extracted. This technique was exceptionally helpful for interpreting the mass scans in cases, when a significant amount of residual gas contamination was still evident in the vacuum system. To illustrate how well this technique works, corresponding mass scans are shown for the mass region from 74 u to 92 u in fig. \ref{sourceoffsetlabel}. The scans were performed shortly after a He bottle change, when still a significant amount of residual gas contamination (most likely hydrocarbons) was evident. Besides the $^{229}$Th$^{3+}$ and $^{233}$U$^{3+}$ signals, two further signals are unambiguously shown to originate from the source close to mass unit 90. A significant rise of these signals was detected always when traces of air had entered the stopping cell (e.g. via the gas feeding line after a change of the He bottle), in parallel a reduction of the $^{229}$Th$^{3+}$ and $^{233}$U$^{3+}$ signals occured. We therefore conclude that an argon adduct might be allowed for Th$^{3+}$ (also tentatively indicated in fig. \ref{mass_scan}).

As can be inferred, a dominant fraction of the extracted $^{229}$Th recoil ions was found to be in the 3+ charge state. An absolute counting rate of ($500\pm60$)/s at peak maximum was found. About 300 counts per second were found to be in the 2+ charge state, while only about 20 counts per second could be obtained in the 1+ charge state. These values do not take any detection efficiency of the MCP detector into account. Typical values of 40-60 \% ion detection efficiency in the considered energy range of 2 to 6 keV, depending on the charge state, have been reported \cite{Oberheide}. In addition to thorium, also uranium was extracted in the 1+, 2+ and 3+ charge state, presumably being sputtered off the source by the escaping $\alpha$ particles. Hydrogen, as well as oxygen, molecule formation is found to occur despite of the high vacuum cleanliness. Furthermore, the whole $^{233}$U decay chain appears in the $2+$ charge state. There are 4 further reproducible signals contained in the scan (as indicated by the double question marks in fig. \ref{mass_scan}), which were proven to originate from the source, but have not yet been identified. The 3 lines at m/q values of 79 u/e, 118.5 u/e and 237 u/e clearly correspond to the same mass of 237 u in different charge states. One further line is seen at m/q $\approx 56$ u/e.

\subsection{Silicon detector measurements}
In order to validate the high $^{229}$Th extraction rate in the 3+ charge state, a second measurement was performed, based on the subsequent $\alpha$ decay of $^{229}$Th. This time a silicon detector (Ametek type BU-016-300-100, thickness 100 $\mu$m, active area 300 mm$^{2}$) was placed in the beam axis at a distance of about 25 mm to the extraction-electrode exit (7) in Fig. 1. Furthermore, an MCP detector was placed sideways under an angle of 45° to the electrode exit. The voltage of -2 kV, applied to the MCP detector surface, was high enough to attract and detect the ions, even though the detector was not placed along the beam axis. In this way, mass scans could be performed and the QMS was set to extract $^{229}$Th$^{3+}$ with a mass resolution of $\Delta$m$/q = 1$ u/e. It should be mentioned that, with this mass resolution, the daughter nuclides of $^{229}$Th could be suppressed even in the 3+ charge state as having a mass distance of $m/q = 4/3$ u/e, while $^{228}$Th$^{3+}$ from the $^{232}$U decay chain was not significantly suppressed. When the QMS had been adjusted, the MCP voltage was taken down and instead a voltage offset of -1250 V was applied to the silicon detector surface to attract the ions. In order to allow for this voltage offset, the silicon detector was electrically isolated from its surrounding and the bias was short-circuited. In this way, the recoil ions were exclusively collected on the silicon detector surface for 5 days. From time to time mass scans were performed and the QMS voltages were adjusted if required. After 5 days of accumulation, the silicon detector was taken out of the setup and placed in a different vacuum chamber. The $\alpha$ decays, subsequently occuring on its surface, were registered and the resulting $\alpha$ energy spectrum was measured for 100 days of continuous detection (with a typical total decay rate of $8.3 \cdot 10^{-3}$/s).\\

\begin{figure*}
 \begin{center}
 \resizebox{0.75\textwidth}{!}{%
 \includegraphics[totalheight=9cm]{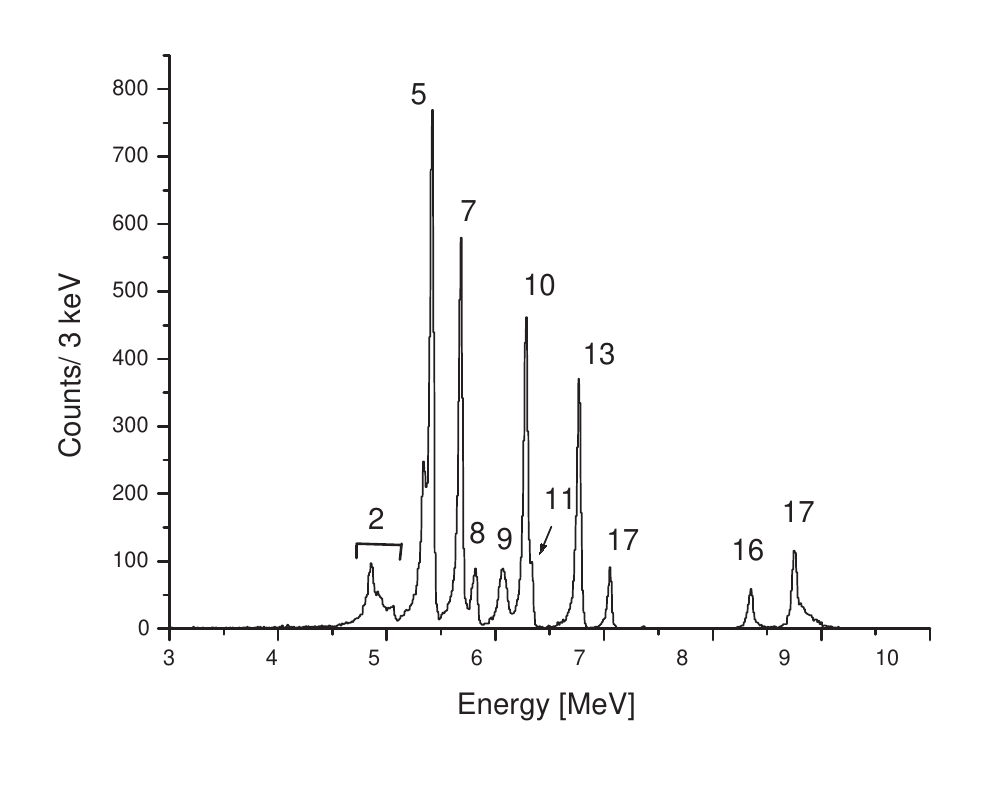}}
 \caption{$\alpha$-energy spectrum as obtained after 5 days of continuous recoil ion accumulation on the surface of a silicon detector and further 100 days of decay measurement. The energy resolution is about 30 keV. The line assignments are listed in table \ref{alphaassign}. For accumulation, the mass spectrometer was set to extract $^{229}$Th$^{3+}$ with a mass resolution of $\Delta$m/q=1 u/e. $^{229}$Th and $^{228}$Th were accumulated with about the same efficiency. The line broadening of the $^{229}$Th $\alpha$ decay arises from
the fact that only 56\% of all decays occur with the dominant energy of 4.845 MeV. Lines of all daughter nuclides are contained in the spectrum, as their half lives are significantly shorter than t$_{1/2}$=7932 yr of $^{229}$Th.}
 \label{Si_spectrum}
 \end{center}
\end{figure*}

The corresponding $\alpha$-energy spectrum is shown in fig. \ref{Si_spectrum}. In order to interpret this spectrum, it should be noted that today's fraction of $^{232}$U, contained as trace contamination in the $^{233}$U source is $N_{232}/N_{233}=(3.9\pm0.4)\cdot 10^{-7}$. Assuming the same extraction efficiencies for $^{229}$Th and $^{228}$Th, the expected relative activity in the $\alpha$ spectrum is estimated to be
\begin{equation}
\resizebox{0.5\textwidth}{!}{$%
\begin{aligned}
\frac{A_{^{228}\text{Th}}}{A_{^{229}\text{Th}}}&=\frac{\lambda_{^{232}\text{U}}\cdot \lambda_{^{228}\text{Th}}\cdot N_{^{232}\text{U}}\cdot F_{^{228}\text{Th}}}{\lambda_{^{233}\text{U}}\cdot \lambda_{^{229}\text{Th}}\cdot N_{^{233}\text{U}}\cdot F_{^{229}\text{Th}}}\\
&=3.9\pm0.4.
\end{aligned}$}
\end{equation}
Therefore, even though the fraction of $^{232}$U in the source is small, the $\alpha$ spectrum is dominated by the decay chain of $^{228}$Th. The relative activity derived from the $\alpha$ spectrum in fig. \ref{Si_spectrum} amounts to $A_{^{228}\text{Th}}/A_{^{229}\text{Th}}=3.6\pm0.7$, and is therefore well in agreement with the expected value. The absolute number of detected $^{229}$Th decays is measured to be $54\pm5$ per day. The uncertainty is dominated by the choice of the integration limits and a possible systematic background from a line overlap with $^{228}$Th and therefore was estimated to be 10\%. Under the valid assumption that 50\% of all $\alpha$ particles are detected, as being emitted into the hemisphere of the silicon detector, the actually accumulated $^{229}$Th activity is $(1.25\pm0.1)\cdot 10^{-3}$ s$^{-1}$. This leads to an absolute number of $(4.5\pm0.3) \cdot 10^{8}$ $^{229}$Th ions collected within 5 days. The absolute $^{229}$Th$^{3+}$extraction rate is then determined to be $(1.0\pm0.1)\cdot 10^3$ s$^{-1}$, which is even significantly higher than the rate obtained from the MCP based measurements. We attribute this difference to the detection efficiency of the MCP detector of about 50 \%. The extraction and purification efficiency for $^{229}$Th$^{3+}$ is accordingly determined to be $(10\pm2)$\%.

Comparable measurements were performed to also determine the extraction efficiencies for $\alpha$-recoil ions originating from isotopes contained in the decay chains from $^{229}$Th and $^{228}$Th downwards. The only differences to the measurement described above were that, due to the shorter half-lives of the considered isotopes, the measurement was performed in parallel to ion accumulation on the silicon detector. Also the measurement times were shorter (typically 2 hours). Especially for the detection of the very short-lived isotopes like $^{217}$At and $^{216}$Po, it was not possible to separate the accumulation and detection. In order to make the measurements comparable, the whole detection system was set onto a voltage offset of -1250 V. The obtained results are listed in table \ref{efficiency} for all extracted charge states. The values listed for Th$^{1+}$ and Th$^{2+}$ are obtained from the MCP-based measurement, taking into account the MCP detection efficiency of about 50\%. Only one extraction efficiency value for Radium is listed, as this measurement involves the relatively long half lives of 14.9 d ($^{225}$Ra), 10.0 d ($^{225}$Ac) and 3.6 d ($^{224}$Ra). This makes the measurements for Ra in all charge states a lengthy procedure, which was not performed. There is no $\alpha$ decay which would allow for the determination of the extraction efficiency for Tl.

\begin{table*}
\begin{footnotesize}
\begin{center}
\begin{tabular}{|c|c|c|c|c|}
\hline
Element& $q=1+$ [\%] & $q=2+$ [\%] & $q=3+$ [\%]& total [\%]\\
\hline
Thorium & $0.34\pm0.07^*$ & $5.5\pm1.1^*$ & $10\pm2.0$ & $16\pm3.2$\\
\hline
Radium$^{*}$ & n.a. & $36\pm7.2$ & n.a. & n.a.\\
\hline
Francium & $21\pm4.2$ & $16\pm3.2$ & $\le1.6\cdot10^{-3}$ & $37\pm7.4$\\
\hline
Radon & $5.8\pm1.2$ & $9.3\pm1.9$ & $(5.3\pm1.1)\cdot10^{-2}$ & $15\pm3.0$ \\
\hline
Astatine & $8.6\pm1.7$ & $13\pm2.6$ & $(3.3\pm0.7)\cdot10^{-2}$ & $22\pm4.4$\\
\hline
Polonium & $7.3\pm1.5$ & $8.1\pm1.6$ & $\le2.1\cdot10^{-3}$ & $15\pm3.0$\\
\hline
Bismuth & $4.3\pm0.9$ & $21\pm4.2$ & $(8.3\pm1.6)\cdot10^{-2}$ & $25\pm5.0$\\
\hline
Lead & $2.2\pm0.4$ & $11\pm2.2$ & $\le1.2\cdot10^{-2}$ & $13\pm2.6$\\
\hline
\end{tabular}
\end{center}
\caption{Ion extraction efficiencies of the ions contained in both ($^{233}$U and $^{232}$U) decay chains, as obtained from the silicon detector measurements. The values are listed separately for the 1+, 2+ and 3+ charge states. For data marked with an asterisk, see explanation in the main text.}
\label{efficiency}
\end{footnotesize}
\end{table*}

\section{Discussion and perspectives}
For the combined extraction and purification efficiency of $^{229}$Th$^{3+}$ from the MLL buffer-gas stopping cell, a value of $(10\pm2)$\% is obtained. This value is supported by two independent measurements, first by ion counting with an MCP detector, second from the successive $\alpha$ decay of $^{229}$Th implanted onto the surface of a silicon detector. To our knowledge, this is the first time that a dominant fraction of $\alpha$-recoil ions has been extracted in the $3+$ charge state from a buffer-gas stopping cell. Usually, the dominant fraction of recoil ions is found to be in the 1+ or 2+ charge state. Indeed, this is also the case for all other nuclides in the $^{233}$U decay chain except for thorium. We attribute this finding to the small 3+ ionization energy of thorium of just 20.0 eV and the high cleanliness of the extraction system as well as the fast extraction times in the ms range.

 When the originally highly charged $\alpha$-recoil ions are stopped in helium buffer gas, electrons are captured until the remaining ionization energy of the recoil ion is smaller than the ionization energy of helium (24.6 eV). In this case, it is energetically favourable for the electron to stay attached to the helium atom. However, even small amounts of impurities in the helium gas lead to further charge-state reduction of $\alpha$-recoil ions. Therefore, only the combination of the exceptionally small 3+ ionization energy of thorium and the high cleanliness of the system did allow for the high 3+ extraction efficiency. For comparison, the 1+, 2+ and 3+ ionization energies of all atoms, which are contained in the decay chains, are listed in table \ref{ionization} \cite{ionize}.
\begin{table*}
\begin{footnotesize}
\begin{center}
\begin{tabular}{|c|c|c|c|}
\hline
Element&1+ [eV] &2+ [eV] &3+ [eV]\\
\hline
Uranium &6.1 & 11.6 & 19.8 \\
\hline
Thorium &6.3 & 11.9 & 18.3\\
\hline
Radium & 5.3 & 10.1 & 31.0\\
\hline
Francium & 4.1 & 22.4 & 33.5\\
\hline
Radon & 10.7 & 21.4 & 29.4\\
\hline
Astatine & 9.3 & 17.9 & 26.6\\
\hline
Polonium & 8.4 & 19.3 & 27.3\\
\hline
Bismuth & 7.3 & 16.7 & 25.6\\
\hline
Lead & 7.4 & 15.0 & 31.9\\
\hline
Thallium & 6.1 & 20.4 & 29.8\\
\hline
\end{tabular}
\end{center}
\caption{Ionization energies for the first 3 charge states of ions contained in the $^{233}$U and $^{232}$U decay chains. Only thorium and uranium exhibit 3+ ionization potentials which are below the 1+ ionization potential of helium (24.6 eV).}
\label{ionization}
\end{footnotesize}
\end{table*}
It is inferred that in this measurement only thorium and uranium can be significantly extracted in the 3+ charge state, which is in agreement with the obtained extraction efficiencies.

Detailed studies of molecule formation rates in a buffer-gas cell were carried out in \cite{Kudryavtsev}. However, only one value for thorium can be found, which is the rate constant for the reaction Th$^+$+O$_2$. The corresponding value of $k=6.0\cdot 10^{-10}$ cm$^3$/s \cite{Johnsen} is the largest listed reaction rate. Thus thorium can be considered as a highly reactive element, which makes a highly purified stopping gas as well as a short ion extraction time prerequisites for an efficient thorium ion extraction. The stopping of thorium ions occurs within 40 mbar of He, which is catalytically purified, so that remaining gas contaminants reach the sub-ppb level (1 ppb is assumed for the following calculation). In this case, the contaminant molecule density is estimated to be $\rho\approx 1\cdot 10^{9}$ 1/cm$^{3}$. When the number of ions in a considered charge state is denoted with $n$, the depopulation rate of this state by molecule formation is generally described by the differential equation $dn(t)/dt=-k\rho n(t)$ \cite{Kudryavtsev}. In this way, the depopulation time constant reads $\tau=1/k\rho$, which in the considered case takes the value 1.7 s$^{-1}$. This is an approximation for the timescale, in which molecule formation is expected to play a significant role. As the extraction time in the ms range is about 3 orders of magnitude shorter, molecule formation is expected to be suppressed, which is in agreement with the obtained measurements.   
 
  The extraction of $^{229}$Th$^{3+}$ is of special importance, as it is the charge state which allows for a simple laser-cooling scheme, as will later be required for laser spectroscopy of $^{229m}$Th \cite{Campbell}. Therefore, the direct extraction of the 3+ charge state will significantly simplify any future approach of ion cooling. Moreover, to a large amount a purification of $^{229}$Th can already be obtained by charge state separation. Taking into account also the mass resolving power of the QMS and a possible radio chemical purification of the $^{233}$U source material, we state that a Th$^{3+}$ ion beam with practically no contamination by short lived daughter isotopes can be obtained.
 
The absolute extraction rate for $^{229}$Th$^{3+}$ from our presently used $^{233}$U source is found to be $(1.0\pm0.1)\cdot 10^{3}$ s$^{-1}$ after mass purification and behind the extraction electrodes. This experimental finding allows for an up-scaling of our earlier estimates for the absolute counting rate and achievable signal contrast of a possible photonic decay of the isomeric ground-state transition of $^{229m}$Th \cite{Wense}. Assuming 2\% of all $^{229}$Th recoil ions to be in the isomeric state, and assuming further loss factors due to ion collection, the UV optical system and the MCP detection efficiency to be 0.4, 0.2 and 0.1, respectively, the achievable UV photon counting rate is $\sim 0.16$ s$^{-1}$ for our present $^{233}$U source strength. By optical calculations, an achievable focal spot size of 70 $\mu$m diameter was predicted. This leads to a signal to noise ratio of $\sim$ 830:1 for the currently used source, when a typical MCP dark count rate of 0.05 s$^{-1}$mm$^{-2}$ is assumed. There is the possibility to increase both values by a factor of about 5, when using the previously described thinner uranium source with larger area. We finally conclude that the described experimental approach offers a realistic chance to detect the VUV fluorescence photons from the ground-state decay of $^{229m}$Th, even in case that a significant non-radiative decay branch exists. Further, the availability of a relatively strong and purified $^{229}$Th ion beam opens the way to address other possible $^{229m}$Th decay branches, like internal conversion and phononic coupling to the carrier material.\\[1cm]

\begin{acknowledgement}
We acknowledge fruitful discussions with T. Udem, T. Lamour, A. Ozawa, E. Peters, T.W. Hänsch, K. Wimmer, D. Habs, P. Hilz, D. Kiefer, J. Schreiber,  J. Burke, E. Haettner, C. Hornung, M.P. Reiter, J. Ebert, W. Plass, K. Eberhardt, C.E. D\"ullmann, S. Stellmer, G.A. Kazakov, T. Schumm, E. Peik, A. Pálffy, A. Fleischmann, C. Enss, V. Sonnenschein and I.D. Moore. This work was supported by DFG Cluster of Excellence Munich-Centre for Advanced Photonics (MAP) and by DFG (Th956/3-1).
\end{acknowledgement}


\begin{thebibliography}{}
%
%
\bibitem[1]{Peik} E. Peik and C. Tamm, Eur. Phys. Lett. \textbf{61}, (2003) 181.
\bibitem[2]{Kroger_Reich} L.A. Kroger and C.W. Reich, Nucl. Phys. \textbf{A 259}, (1976) 29.
\bibitem[3]{Helmer_Reich} R. Helmer and C.W. Reich, Phys. Rev. \textbf{C 49}, (1994) 1845.
\bibitem[4]{Beck} B.R. Beck {\it et al.}, Phys. Rev. Lett. \textbf{109}, (2007) 142501.
\bibitem[5]{Trzhaskovskaya1} F.F. Karpeshin and M.B. Trzhaskovskaya, Phys. Rev. C \textbf{76}, (2007) 054313.
\bibitem[6]{Trzhaskovskaya2} F.F. Karpeshin and M.B. Trzhaskovskaya, Physics of Atomic Nuclei \textbf{69-4}, (2006) 571. 
\bibitem[7]{Tkalya1} E.V. Tkalya {\it et al.}, Phys. Rev. C \textbf{61}, (2000) 064308.
\bibitem[8]{Tkalya2} E.V. Tkalya, JETP Letters \textbf{71-8}, (2000) 311.
\bibitem[9]{Peik2} E. Peik {\it et al.}, proceedings of $7^{\text{th}}$ symposium on frequency standards and metrology, Pacific Grove CA U.S.A. October 5-11 2008, Frequency Standards and Metrology ,(2009) 532.
\bibitem[10]{Swanberg} E. Swanberg {\it et al.}, Am. Phys. Soc., (2011) F1.002.
\bibitem[11]{Rellergert} W.G. Rellergert {\it et al.}, Phys. Rev. Lett. \textbf{104}, (2010) 200802.
\bibitem[12]{Porsev} S. Porsev {\it et al.}, Phys. Rev. Lett. \textbf{105}, (2010) 182501.
\bibitem[13]{Kazakov} G.A. Kazakov {\it et al.},  New J. Phys.\textbf{14}, (2012) 083019.
\bibitem[14]{Raeder} S. Raeder {\it et al.}, J. Phys. \textbf{B 44}, (2011) 165005.
\bibitem[15]{Zhao} X. Zhao {\it et al.}, Phys. Rev. Lett. \textbf{109}, (2012) 160801.
\bibitem[16]{Peik3} E. Peik and K. Zimmermann, Phys. Rev. Lett. \textbf{111}, (2013) 018901.
\bibitem[17]{Zimmermann} K. Zimmermann, Ph.D. thesis, Univ. Hannover, Germany (2010). 
\bibitem[18]{Swanberg2} E. Swanberg, Ph.D. thesis, University of California, Berkeley (2012).
\bibitem[19]{Wense} L. v.d.Wense {\it et al.}, JINST \textbf{8}, (2013) P03005.
\bibitem[20]{Neumayr} J.B. Neumayr {\it et al.}, Rev. Sci. Instr. \textbf{77}, (2006) 065109.
\bibitem[21]{Ziegler} J.F. Ziegler, version TRIM-2012.03 was used.
\bibitem[22]{Neumayr2} J.B. Neumayr, Ph.D. thesis, Ludwig-Maximilians-Universit\"at M\"unchen, Munich, Germany (2004).
\bibitem[23]{Brubaker} W.M. Brubaker, Advances in mass spectrometry \textbf{4}, (1968) 293.
\bibitem[24]{Haettner} E. Haettner, Ph.D. thesis, Univ. Giessen, Germany (2011).
\bibitem[25]{ZBL} J.F. Ziegler, J.P. Biersack, U. Littmark, \textit{The Stopping and Range of Ions in Matter} (Pergamon, New York 1985).
\bibitem[26]{mdh} K. Nordlund, Comput. Mater. Sci. 3,(1995), 448.
\bibitem[27]{Hashimoto} T. Hashimoto {\it et al.}, J. inorg. nucl. Chem. \textbf{43-10}, (1981) 2233.
\bibitem[28]{Laval} S.E. Taylor {\it et al.}, Phys. Chem. Chem. Phys. \textbf{10}, (2008) 422.
\bibitem[29]{Matzke} H. Matzke, Radiation Effects \textbf{53}, (1980) 219.
\bibitem[30]{Fick} A. Fick, Phil. Mag. \textbf{10}, (1855) 30.
\bibitem[31]{Sonnenschein} V. Sonnenschein {\it et al.}, Eur. Phys. J. \textbf{A 48}, (2012) 52. 
\bibitem[32]{Bateman} H. Bateman, Proc. Cambridge Phil. Soc. IS \textbf{423}, (1910).
\bibitem[33]{Magill} J. Magill, J. Galy, \textit{Radioactivity Radionuclides Radiation} (Springer-Verlag, Berlin Heidelberg 2005) 47.
\bibitem[34]{Oberheide} J. Oberheide {\it et al.}, Meas. Sci. Technol. \textbf{8}, (1997) 351.
\bibitem[35]{gov} NNDC Interactive Chart of Nuclides [Online].
Available {\verb http://www.nndc.bnl.gov/chart } [2014, November 26]. Brookhaven National Laboratory, Brookhaven.
\bibitem[36]{ionize} A. Kramida, Yu. Ralchenko, J. Reader {\it et al.}, NIST Atomic Spectra Database (ver. 5.2), [Online]. Available {\verb http://physics.nist.gov/asd } [2014, November 26]. National Institute of Standards and Technology, Gaithersburg, MD.
\bibitem[37]{Kudryavtsev} Yu. Kudryavtsev  {\it et al.}, Nucl. Instr. and Meth. in Phys. Res. B \textbf{179}, (2001) 412
\bibitem[38]{Johnsen} R. Johnsen {\it et al.}, J. Phys. Chem. \textbf{61}, (1974) 5404.
\bibitem[39]{Campbell} C.J. Campbell {\it et al.}, Phys. Rev. Lett. \textbf{106}, (2011) 223001.

\end{thebibliography}
%

\end{document}